\documentclass[lettersize,journal]{IEEEtran}

\usepackage{amssymb,amsmath}
\usepackage{cite}
\usepackage{graphicx}
\usepackage{subfig}
\usepackage{psfrag}
\usepackage{url}
\usepackage[utf8]{inputenc}
\usepackage[absolute,overlay]{textpos}
\usepackage{gensymb}
\usepackage{cases}
\usepackage[font=footnotesize]{caption}
\usepackage{float}
\usepackage[linesnumbered,ruled,lined]{algorithm2e}
\usepackage{pgf}
\usepackage[nocomma]{optidef}
\usepackage{amsthm}
\usepackage{booktabs}
\usepackage{color,soul}
\usepackage{textcomp}
\usepackage{bbm}
\usepackage{xcolor}
\usepackage{tabularx}
\usepackage{makecell}
\usepackage{multirow}
\usepackage{stfloats}
\usepackage{placeins}

\renewcommand{\arraystretch}{0.95}

\DeclareMathOperator{\Tr}{Tr}

\def\BibTeX{{\rm B\kern-.05em{\sc i\kern-.025em b}\kern-.08em
    T\kern-.1667em\lower.7ex\hbox{E}\kern-.125emX}}

\begin{document}

\title{FlowSem: Flow Matching for Adaptive Wireless Image Transmission in Semantic Communication}

\author{
\IEEEauthorblockN{Arnab Emon and Eslam Eldeeb,~\IEEEmembership{Member,~IEEE} }
\thanks{The authors are with the Centre for Wireless Communications (CWC), University of Oulu, Finland. (e-mail: eslam.eldeeb@oulu.fi; arnab.emon@oulu.fi;).}

    \thanks{This work was supported by the Business Finland project, 6G-FISRE.}
}

\maketitle

\begin{abstract}
Wireless image transmission becomes challenging under poor channel conditions and stringent bandwidth constraints, when the receiver needs to preserve both pixel-level fidelity and meaningful visual structure. Classical separation-based systems, such as better portable graphics with low-density parity-check coding (BPG+LDPC), may suffer from cliff-effect behavior. While deep joint source-channel coding (DeepJSCC) provides graceful degradation as channel conditions worsen, its reconstructions may still lose fine image details under strong compression and severe channel distortion. To address this limitation, this paper proposes a two-stage flow matching-based semantic communication framework, termed FlowSem, for wireless image transmission. In the first stage, a signal-to-noise ratio (SNR)-adaptive DeepJSCC model maps the source image into channel symbols and produces a coarse reconstruction at the receiver. In the second stage, a conditional flow matching model generates the final image from Gaussian noise while being conditioned on the DeepJSCC reconstruction and the channel SNR. The proposed framework is evaluated on the Cityscapes dataset under additive white Gaussian noise (AWGN) and Rayleigh fading channels using a fixed channel-symbol budget. Several baselines are considered including a rate-matched BPG+LDPC system, DeepJSCC, and denoising diffusion probabilistic model (DDPM) and denoising diffusion implicit model (DDIM). Simulation results show that the proposed FlowSem achieves competitive pixel-level fidelity and improved structural and perceptual reconstruction quality over the considered generative baselines across different channel conditions. FlowSem demonstrates up to $60 \%$ lower Fr\'echet inception distance (FID) compared to the diffusion baselines in low SNR ranges. Moreover, FlowSem reaches high reconstruction quality using only a few ODE integration steps and provides a favorable quality–latency tradeoff compared with both standard DDPM sampling and accelerated DDIM sampling. Source codes are available through~\url{https://github.com/ArnabEmon24/FlowSem}.
\end{abstract}

\begin{IEEEkeywords}
Deep joint source-channel coding, diffusion models, flow matching, generative reconstruction, semantic communication, wireless image transmission.
\end{IEEEkeywords}

\section{Introduction}
\label{sec:intro}
Semantic communication is expected to be a key enabler in next-generation communication systems. While conventional communication systems primarily target reliable transmission and recovery of information symbols, semantic communication places greater emphasis on delivering relevant information to the intended task or meaning at the receiver~\cite{qin2021semantic,gunduz2023beyond}. The received semantic information can be used to either reconstruct the original data or to recover a task-relevant representation that is sufficient to achieve a certain objective. Semantic communication demonstrates several advantages, such as reducing bandwidth usage and latency while maintaining efficient data reconstruction or task execution~\cite{10948463}.

Wireless image transmission plays an increasingly important role in many modern applications such as autonomous driving, remote sensing, smart-city monitoring, surveillance, and augmented and virtual reality. In these applications, the transmitted image may be used either for human interpretation or for downstream tasks, including  scene understanding, object detection, and navigation. From a semantic communication perspective, the objective is therefore not only to preserve pixel-level fidelity, but also to retain meaningful visual structures while operating under limited bandwidth and communication resources.

Conventional wireless image transmission generally follows the source-channel separation principle, where source coding, \emph{e.g.}, better portable graphics (BPG)~\cite{bellard2018bpg}, and channel coding, \emph{e.g.}, low-density parity-check (LDPC) coding~\cite{gallager1962ldpc}, are designed as separate components. The source coder compresses the image into a bitstream, while the channel coder introduces structured redundancy to protect the transmitted stream against channel errors~\cite{xu2022adjscc,wang2022perceptual}. Nevertheless, practical digital image transmission may exhibit a cliff effect under limited channel resources. In that case, once channel decoding becomes unreliable, small degradation in channel quality can lead to severe deterioration in the reconstructed image~\cite{bourtsoulatze2019deepjscc,10702555}.

Recently, deep-learning-based systems, such as deep joint source-channel coding (DeepJSCC), have shown promising performance for wireless image transmission. DeepJSCC employs a neural autoencoder to directly map the source image into channel symbols and reconstruct the image from the received noisy symbols. Thus, DeepJSCC breaks the explicit intermediate digital bitstream construction~\cite{xie2021deepsc,dai2022ntscc}. However, DeepJSCC remains subject to perception-distortion tradeoff, where minimizing pixel-wise reconstruction distortion does not necessarily lead to the best perceptual quality~\cite{blau2018perception}. In addition, the reconstructed images may lose important information like edge sharpness and fine textures under poor channel conditions and stringent bandwidth.

Recently, generative models have been investigated to improve the perceptual quality of reconstructed images. For instance, denoising diffusion probabilistic models (DDPMs) have demonstrated strong performance in image generation and restoration~\cite{ho2020ddpm}. A DDPM often includes a forward-backward process where, in the forward process, the model gradually corrupts an image into Gaussian noise, while a learned reverse process denoises the noisy image back to a clean image. By learning a rich prior over the image distribution, diffusion models can effectively recover visually plausible details that may be missing from a degraded reconstruction. Consequently, diffusion models have been widely adopted for image restoration and refinement tasks. 

However, an important limitation in diffusion models lies in their iterative sampling process. In standard DDPM, inference requires repeated evaluations over a long denoising trajectory, which can lead to high computational complexity and receiver-side latency. Although accelerated methods, such as denoising diffusion implicit models (DDIMs)~\cite{song2021ddim}, can reduce the number of required steps, the quality-latency tradeoff remains important for practical deployments in wireless receivers.

Flow matching has recently emerged as an alternative generative framework~\cite{lipman2023flowmatching}. Flow matching learns a time-dependent velocity field that transports samples from a simple noise distribution toward the target data distribution along a continuous probability path~\cite{liu2023rectifiedflow}. Sampling is performed by numerically integrating an ODE defined by the learned velocity field. This formulation can provide efficient deterministic sampling using a relatively small number of network evaluations.

In this paper, we propose flow matching-based semantic communication (FlowSem), an approach for wireless semantic image transfer. We propose a two-stage wireless image semantic communication framework, where we develop an adaptive DeepJSCC model in the first stage that encodes and reconstructs the image using a deep convolutional autoencoder. In the second stage, a conditional generative flow matching model is developed to refine the reconstructed image.
The main contributions of this paper are summarized as follows:
\begin{itemize}
    \item We propose FlowSem, a two-stage wireless image semantic communication framework that combines an SNR-adaptive DeepJSCC encoder-decoder with conditional flow matching for wireless image reconstruction. The DeepJSCC stage produces a coarse reconstruction from the received channel symbols, while the flow matching stage performs generative refinement at the receiver.

    \item We develop an SNR-conditioned flow matching model that generates the final reconstructed image conditioned jointly on the coarse DeepJSCC reconstruction and the channel SNR. Thus, this allows a single trained model to operate over a wide range of SNR conditions.

    \item We evaluate FlowSem over additive white Gaussian noise (AWGN) and Rayleigh fading channels while comparing it with recent baselines, including rate-matched BPG+LDPC and DeepJSCC. In addition, we consider conventional diffusion-based semantic communication (DDPM) and accelerated diffusion-based semantic communication (DDIM) as generative-based baselines. 

    \item We evaluate FlowSem compared with the considered baselines in terms of distortion-based, structural, perceptual, and distribution-level metrics. Moreover, we investigate the quality-complexity tradeoff of generative reconstruction by studying the number of sampling steps on inference time and reconstruction quality.



    
    
    \item Simulation results demonstrate that FlowSem provides competitive pixel-level reconstruction while consistently improving structural, perceptual, and distribution-level quality over the considered generative baselines. In addition, FlowSem provides a favorable quality–latency tradeoff relative to diffusion-based reconstruction.
    
\end{itemize}

The remainder of this paper is organized as follows. Section~\ref{sec:related} reviews the related literature. Section~\ref{sec:system} presents the system model, while Section~\ref{sec:background} introduces the necessary background on DeepJSCC and diffusion models. Section~\ref{sec:method} presents the proposed FlowSem framework. Section~\ref{sec:results} discusses the experimental results, and Section~\ref{sec:conclusion} concludes the paper.

\section{Related Work}
\label{sec:related}
In this section, we present the recent state-of-the-art literature on semantic communication and wireless image transfer. One of the earliest DeepJSCC frameworks was introduced in~\cite{bourtsoulatze2019deepjscc} as an end-to-end neural image transmission system that directly maps image samples into complex channel symbols. By including the wireless channel as a non-trainable layer during training, the encoder and decoder can jointly learn a representation that is both compact and robust to channel noise. Unlike traditional digital systems, DeepJSCC demonstrates robust performance under diverse channel conditions.


Several works have extended the basic form of DeepJSCC to enable adaptive deployments. For instance, the work in~\cite{xu2022adjscc} proposes attention DeepJSCC (ADJSCC), which introduces an attention feature module into the encoder and decoder. The module uses the channel SNR to adaptively reweigh feature channels during transmission. The authors show that a single ADJSCC model can operate across a wide range of SNR conditions, instead of training a separate model for each channel condition. A similar type of SNR-adaptive design is studied in~\cite{ding2021snradaptive}, where the encoder and decoder are conditioned on the channel SNR to cover a wide operating range with a single model. 

In addition, there are other works that have extended the traditional DeepJSCC to support bandwidth and rate adaptation. These include successive refinement~\cite{kurka2019successive}, layered bandwidth-adaptive transmission~\cite{bian2023deepjscclpp}, feedback-assisted coding~\cite{kurka2020deepjsccf}, OFDM-based transmission over multipath channels~\cite{yang2021ofdm}, adaptive rate control~\cite{yang2022adaptivejscc}, and a MIMO extension~\cite{wu2023deepjsccmimo}. These methods improve transmission flexibility. However, they optimize distortion-based objectives rather than targeting image quality and perception under limited channel resources.

Recently, generative models have also been widely adopted in wireless image transmission targeting perceptual quality enhancement. The work in~\cite{erdemir2023generativejscc} proposes a GAN-based approach for wireless image transmission, where a StyleGAN-2 prior is used to reconstruct the image from the received symbols. The authors show that their approach outperforms DeepJSCC and distortion-based methods in terms of perceptual quality, especially under low bandwidth and low SNR conditions. However, GAN-based methods are often difficult to train. Also, their reconstruction quality depends mostly on data distribution learned by the generative model. 

Diffusion models have recently proved effective for image restoration and inverse problems~\cite{liu2023i2sb}. The work in~\cite{wu2024cddm} introduces a diffusion-based channel denoising model that uses the similarity between the diffusion forward process and wireless channel noise to mitigate channel distortion. However, its denoiser network is designed to reduce channel noise rather than improve perceptual quality. In~\cite{jiang2024diffsc}, a diffusion denoiser is combined with DeepJSCC decoder that uses the decoded image as conditioning information for generation. Their method improves pixel-level reconstruction as well as perception quality. However, their main limitation lies in the large sampling time needed to denoise the received image. Other recent works adopt diffusion models in semantic communication, such as~\cite{zhang2025semanticsguided} that directly applies diffusion models to the wireless image reconstruction problem and~\cite{yilmaz2024perceptual} that studies a diffusion model refinement to the components of a DeepJSCC that lie outside the recovered subspace.

Flow matching provides an alternative generative approach to diffusion models. Instead of learning an iterative reverse Markov denoising process, flow matching learns a continuous velocity field between source and target~\cite{lipman2023flowmatching}. Flow matching has recently been adopted for image restoration problems such as super-resolution, deblurring and image dehazing~\cite{martin2025pnpflow,ray2026hazematching}.
More recently, flow matching has also been investigated for semantic communication. In~\cite{11604027}, the authors propose a land-then-transport algorithm using flow matching for wireless image transmission by incorporating the wireless channel into a flow model at the receiver. Additionally, the work in~\cite{gao2026lowlatencygenerativesemanticcommunication} formulates receiver-side recovery using channel-realization flow matching for semantic communication.

Different from recent flow matching approaches, FlowSem follows a two-stage architecture. First, an SNR-adaptive DeepJSCC transmits the image and produces a coarse reconstruction. Then, a conditional flow matching model refines the received image conditioned on both the DeepJSCC output and channel SNR.

\section{System Model}
\label{sec:system}
Consider a point-to-point wireless image transmission system where a source image $\mathbf{s} \in [0,1]^{H\times W\times 3}$ is transmitted from a single transmitter to a single receiver over a noisy wireless channel. The transmitter encodes the image into a compact set of channel symbols, while the receiver reconstructs an image $\hat{\mathbf{s}}$ from the received signal. The objective is to preserve both the pixel-level information and the meaningful visual structure of the source image while operating under limited channel symbol budget.



\subsection{Transmitter}

The transmitter maps the source image $\mathbf{s}$ to a compact channel representation using a semantic encoder $f_{\phi}$ as follows:
\begin{equation}
    \mathbf{x} = f_{\phi}(\mathbf{s}, \gamma),
\end{equation}
where $\phi$ denotes the trainable encoder parameters and $\gamma$ is the SNR. To demonstrate the effectiveness of generative modeling in wireless image semantic communication, we consider an adaptive SNR encoder-decoder design, where we assume that the operating SNR $\gamma$ is available at both transmitter and receiver~\cite{ding2021snradaptive}. During encoding, the source image is compressed into a compact latent representation whose size is determined by the target compression ratio. While this representation saves communication resources and reduces transmission latency, it creates a stronger transmission bottleneck that makes accurate reconstruction more challenging. 

\subsection{Wireless Channel}
The transmitted channel symbols are received over the wireless channel as:
\begin{equation}
\label{channel_eq}
    \mathbf{y}=\mathbf{h} \: \odot \: \mathbf{x}+\mathbf{n},
\end{equation}
where $h$ is the wireless channel modeled using Rayleigh fading, where channel coefficients are independently generated and normalized for the transmitted symbols. $\mathbf{n}\sim \mathcal{N}(0,\sigma^{2}\mathbf{I})$ is an AWGN component with variance $\sigma^{2}=P/\gamma$ , average transmit power $P$, and linear signal-to-noise ratio (SNR) $\gamma=10^{\gamma_{\mathrm{dB}}/10}$. Zero-forcing equalization is applied as follows~\cite{proakis2008digital}:
\begin{equation}
\label{equalization}
    \tilde{y}_i=\frac{y_i}{h_i+\zeta}, \quad \quad i = 1, \dots, k,
\end{equation}
where {$\zeta=10^{-10}$} is a small constant used for numerical stability and $k$ is the number of symbols.



\subsection{Receiver}
At the receiver, we consider a semantic decoder that aims to reconstruct the image as follows:
\begin{equation}
    \hat{\mathbf{s}}
    =
    g_{\theta}(\tilde{\mathbf{y}},\gamma),
\end{equation}
where $\theta$ denotes the trainable decoder parameters. 

\subsection{Problem Definition}
Given a source image $\mathbf{s}$, the objective is to reconstruct an image $\hat{\mathbf{s}}$ at the receiver under a fixed channel-symbol budget and transmit-power constraint while preserving both pixel-level fidelity and perceptual structure under different channel conditions. Unlike prior works that train a separate model for each SNR, the objective in this work is to design an SNR-adaptive framework, where a single model can adapt to different SNR ranges. We address this objective in the proposed FlowSem framework through two successive training stages. First, an SNR-adaptive DeepJSCC model is trained to reconstruct the source image from the received channel symbols. Second, we train a conditional flow matching model to refine the coarse output of the frozen DeepJSCC. Section~\ref{sec:method} introduces the proposed FlowSem approach.

\section{Preliminaries}
\label{sec:background}

This section reviews materials that are essential to present the proposed methodology in Section~\ref{sec:method}. We first overview the architecture of DeepJSCC, then we present the formulations of diffusion models used for the generative baseline. 

\subsection{Deep Joint Source-Channel Coding}
\label{subsec:deepjscc}

DeepJSCC replaces the traditional separate source coding, channel coding and modulation blocks with a single neural autoencoder. The autoencoder is trained end-to-end together over a wireless channel. Although Shannon's separation theorem is theoretically optimal for long block lengths~\cite{shannon1948mathematical}, practical wireless image transmission generally operates with short block lengths and limited channel resources. Therefore, traditional approaches become less effective in practice. DeepJSCC is proposed as an alternative to traditional approaches for wireless image transfer.

In DeepJSCC, the encoder and decoder are usually designed as a deep autoencoder using convolutional layers that are used to extract a compact representation of the image. The DeepJSCC model is trained end-to-end by minimizing the distortion between the source image and the reconstructed image:
\begin{equation}
    \mathcal{L}_{\mathrm{JSCC}}
    =
    \mathbb{E}
    \left[
    \left\|
    \mathbf{s}-\hat{\mathbf{s}}
    \right\|_2^2
    \right].
\end{equation}
DeepJSCC provides an efficient framework for joint source-channel image transmission. However, under stringent budgets and low SNRs, the reconstructed images may lose fine visual details as the decoder is primarily trained via a distortion-based reconstruction objective.

\subsection{Diffusion Models}
\label{subsec:diffusion}

A denoising diffusion probabilistic model (DDPM) is a latent-variable generative model consisting of two Markov processes, namely, forward and backward processes~\cite{ho2020ddpm}. The forward process gradually adds noise to the input image until it becomes nearly pure noise. The reverse process then learns to remove the noise iteratively to reconstruct the original image.

Given an input sample, $\mathbf{x}_0$, the forward process adds Gaussian noise over $T$ steps as follows:
\begin{equation}
    q(\mathbf{x}_t\mid\mathbf{x}_{t-1})
    =
    \mathcal{N}\!\left(
    \mathbf{x}_t;\sqrt{1-\beta_t}\,\mathbf{x}_{t-1},\,\beta_t\mathbf{I}
    \right),
\end{equation}
where $\beta_t$ is the noise variance at step $t$. Its value is determined by a predefined schedule ranging from $\beta_1$ to $\beta_T$. Let $\alpha_t=1-\beta_t$ and $\bar{\alpha}_t=\prod_{\tau=1}^{t}\alpha_{\tau}$. Using these definitions, the marginal noising distribution can be written in the closed form as:
\begin{equation}
    \mathbf{x}_t
    =
    \sqrt{\bar{\alpha}_t}\,\mathbf{x}_0
    +
    \sqrt{1-\bar{\alpha}_t}\,\boldsymbol{\epsilon},
    \label{eq:diff_closedform}
\end{equation}
where $\boldsymbol{\epsilon}\sim\mathcal{N}(0,\mathbf{I})$ is the standard Gaussian vector. This formulation allows any noisy sample $\mathbf{x}_t$ to be generated directly from the clean input $\mathbf{x}_0$ without simulating all intermediate diffusion steps. During the reverse process, it starts from pure Gaussian noise $\mathbf{x}_T\sim\mathcal{N}(0,\mathbf{I})$, and progressively removes the noise to reconstruct the clean input. In its standard formulation, a denoising network $\boldsymbol{\epsilon}_{\psi}$ is trained to predict the Gaussian noise added during the forward process by minimizing the following loss:
\begin{equation}
\label{simple_diff}
    \mathcal{L}_{\mathrm{diff}}
    =
    \mathbb{E}_{\mathbf{x}_0,\boldsymbol{\epsilon},t}
    \left[
    \left\|
    \boldsymbol{\epsilon}
    -
    \boldsymbol{\epsilon}_{\psi}(\mathbf{x}_t,t)
    \right\|_2^2
    \right],
\end{equation}
where $\psi$ denotes the learnable parameters of the denoising network $\boldsymbol{\epsilon}_{\psi}$.

The conventional noise-prediction training objective is shown in~\eqref{simple_diff}. For the conditional diffusion baselines considered in this paper, we train the network to directly predict the clean image using the following reconstruction and conditioning-consistency objective~\cite{kingma2021vdm,karras2022edm,salimans2022progressive}. Therefore, the underlying objective is constructed as follows:
\begin{equation}
\begin{aligned}
    \mathcal{L}_{\mathrm{diff}}
    =
    \mathbb{E}_{\mathbf{x}_0,t,\boldsymbol{\epsilon}}
    \Big[
    &
    \left\|\hat{\mathbf{x}}_0-\mathbf{x}_0\right\|_1
    +
    \delta\left\|\hat{\mathbf{x}}_0-\mathbf{x}_0\right\|_2^2
    \\
    &
    +
    \omega\left\|\hat{\mathbf{x}}_0-\mathbf{c}\right\|_1
    \Big],
\end{aligned}
\end{equation}
where $\delta$ weights the $L_2$ reconstruction term, $\mathbf{c}$ is the output of the semantic decoder (\emph{e.g.}, $\hat{\mathbf{s}}_{\mathrm{JSCC}}$), and $\omega$ weights the consistency term, which encourages the generated image to remain related to the received signal. In addition $\hat{\mathbf{x}}_0$ is estimated as follows:
\begin{equation}
    \hat{\mathbf{x}}_0
    =
    D_{\psi}(\mathbf{x}_t,t,\mathbf{c},\gamma),
\end{equation}
where $D_{\psi}$ is the conditional diffusion network.

DDIM~\cite{song2021ddim} accelerates sampling compared with the conventional DDPM formulation. In particular, it uses the same trained diffusion model but performs inference using a subsampled sequence of diffusion timesteps. By learning the underlying image distribution, diffusion models provide a rich prior that can recover missing visual details. However, standard DDPM sampling requires a long sequence of reverse steps, which can lead to considerable inference time. While DDIM may reduce this cost by skipping intermediate steps, the quality-latency tradeoff remains critical in diffusion models. In the next section, we present flow matching, an alternative to denoising diffusion, that provides high quality generation while relying on relatively faster inference formulations.

\section{Flow Matching-Based Semantic Communication}
\label{sec:method}
In this section, we introduce the proposed flow matching-based semantic communication framework, FlowSem. We propose a two-stage architecture consisting of an SNR-adaptive DeepJSCC stage and a conditional flow matching model refinement stage at the receiver.


\begin{figure*}[t!]
    \centering
    \includegraphics[width=0.90\textwidth,trim={0 0 0 0},clip]{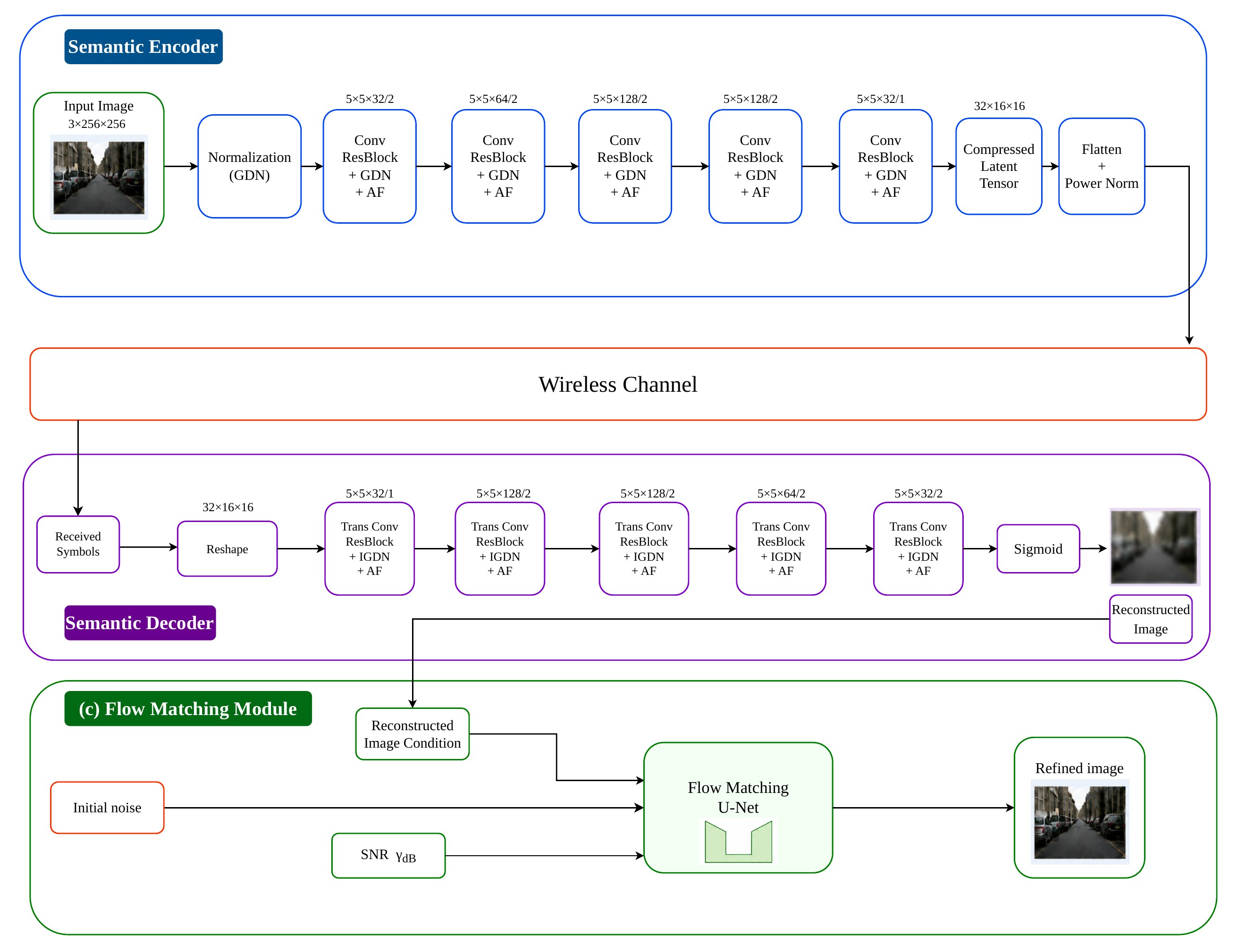}
    \vspace{2mm}
    \caption{End-to-end architecture of FlowSem, consisting of (a) the SNR-adaptive semantic encoder, (b) the SNR-adaptive semantic decoder, and (c) the conditional flow matching refinement module.}
    \vspace{0mm}
    \label{fig:system_model}
\end{figure*}

\subsection{Phase 1: Adaptive DeepJSCC}
Let $\mathbf{s}$ be the source image and $\gamma$ be the channel SNR. We consider an SNR-adaptive DeepJSCC~\cite{xu2022adjscc}, where the encoder $f_{\phi}$ maps the image jointly with the SNR into a latent representation $z$. We construct the adaptive encoder as follows: 
\begin{equation}
\label{deepjscc_encoder}
    \mathbf{x}=f_{\phi}(\mathbf{s},\gamma),
\end{equation}
where the encoder $f_{\phi}$ is modeled as a deep convolutional neural network (CNN) consisting of residual blocks. As shown in Fig.~\ref{fig:system_model}, each residual block consists of convolutional layers followed by a batch normalization layer and a generalized divisive normalization (GDN) layer~\cite{balle2016gdn}. An attention feature (AF) block is inserted after each residual block, which generates channel-wise scaling coefficients to reweigh the feature channels according to the channel quality.

The output of the residual blocks is the latent representation $\mathbf{x}$, which is normalized and flattened into $k$ real-valued channel symbols. Then, the latent representation is transmitted over the wireless channel. The compression ratio is defined as follows:
\begin{equation}
    \mathrm{CR}=\frac{n}{k},
\end{equation}
where $n=3HW$ is the source pixel values per image. $H$ and $W$ denote the image height and width, respectively. A larger $\mathrm{CR}$ corresponds to transmitting fewer channel symbols relative to the number of source-image values, which represents a more challenging communication bottleneck. In the considered experiment, each transmitted channel symbol corresponds to one real-valued encoder output. The received signal $\tilde{\mathbf{y}}$ is obtained as in~\eqref{channel_eq} and~\eqref{equalization}. At the receiver, the adaptive decoder $g_{\theta}$ reconstructs the original image as follows:
\begin{equation}
    \label{jscc_decoder}
    \hat{\mathbf{s}}_{\mathrm{JSCC}}=g_{\theta}(\tilde{\mathbf{y}},\gamma),
\end{equation}
where $\hat{\mathbf{s}}_{\mathrm{JSCC}}$ is the output of the DeepJSCC phase. The decoder mirrors the encoder using transposed convolutional residual blocks that contain transposed convolutional layers, inverse GDN (IGDN) layers, and AF layers. Finally, a sigmoid activation layer is used to keep the reconstructed values within the valid image range.

In the deployed autoencoder framework, each residual block includes a skip connection that adds the block's input to the block's output, following the residual learning principle of~\cite{he2016resnet}. This helps to shorten the path for gradient backpropagation and eases training when network depth increases. Within each block, GDN serves as the nonlinear normalization step. It is chosen because it models the statistics of natural images more effectively than standard activation functions. When a residual block changes the spatial resolution or the number of feature channels, the skip path is adjusted to match the output dimensions before the features are added. An attention feature (AF) block is inserted in both the encoder and the decoder. Each AF block takes the current SNR and an intermediate feature map as inputs and generates channel-wise scaling coefficients. These coefficients adapt intermediate feature representation according to the operating SNR. Consequently, a single encoder-decoder model can be trained over the considered SNR range instead of training a separate model for each SNR value. The DeepJSCC model is trained end-to-end by minimizing the distortion between the source image and the coarse reconstructed image:
\begin{equation}
\label{deepjscc_loss}
    \mathcal{L}_{\mathrm{JSCC}}
    =
    \mathbb{E}_{\mathbf{s},\gamma_{\mathrm{dB}}}
    \left[
    \left\|
    \mathbf{s}-\hat{\mathbf{s}}_{\mathrm{JSCC}}
    \right\|_2^2
    \right].
\end{equation}
After training, the DeepJSCC encoder and decoder are kept frozen for the second phase.

\begin{algorithm}[!t]

\textbf{Input:} Training dataset


\textbf{Phase 1: Train the DeepJSCC model}

\For{\text{each mini-batch of images} $\mathbf{s}$}{

Sample an SNR $\gamma$ from the training SNR range

Calculate the latent variable $\mathbf{x}$ using~\eqref{deepjscc_encoder}

Transmit the normalized channel symbols through the wireless channel and obtain $\hat{\mathbf{s}}_{\mathrm{JSCC}}$ via~\eqref{jscc_decoder}

Update $f_{\phi}$ and $g_{\theta}$ by minimizing $\mathcal{L}_{\mathrm{JSCC}}$ in~\eqref{deepjscc_loss}

}

\textbf{Return:} Trained DeepJSCC autoencoder $f_{\phi}$ and $g_{\theta}$


\textbf{Phase 2: Train the conditional flow matching model}

\For{each mini-batch of images $\mathbf{s}$}{

Sample a random SNR value $\gamma$

Compute the frozen DeepJSCC reconstruction $\hat{\mathbf{s}}_{\mathrm{JSCC}}$ using~\eqref{deepjscc_encoder},~\eqref{jscc_decoder}

Sample $\mathbf{x}_0\sim\mathcal{N}(0,\mathbf{I})$ and $t\sim\mathcal{U}(0,1)$

Compute $\mathbf{x}_t$ using~\eqref{interpolation}


Update $v_{\psi}$ by minimizing $\mathcal{L}_{\mathrm{FM}}$ using~\eqref{fm_loss}

}

\textbf{Return:} Trained velocity field $v_{\psi}$

\caption{The proposed flow matching-based semantic communication (FlowSem) algorithm.}
\label{alg:overall}
\vspace{0mm}

\end{algorithm}

\subsection{Phase 2: Adaptive Conditional Flow Matching}

Flow matching learns a continuous-time generative process as an alternative to the discrete denoising chain of diffusion models~\cite{lipman2023flowmatching}. Instead of a step-by-step denoiser, flow matching learns a time-dependent velocity field $v_{\psi}(\cdot)$ that defines a probability path from a simple noise distribution $p_0=\mathcal{N}(0,\mathbf{I})$ at $t=0$ to the target data distribution $p_1$ (\emph{e.g.}, image distribution) at $t=1$. In FlowSem, the velocity field is conditioned on the DeepJSCC reconstruction $\hat{\mathbf{s}}_{\mathrm{JSCC}}$ and the channel SNR $\gamma$.

The continuous dynamics are defined through the ODE as follows~\cite{lipman2023flowmatching}:
\begin{equation}
\label{ode_equation}
\frac{d\mathbf{x}(t)}{dt} = v_{\psi}(\mathbf{x}(t),t, \hat{\mathbf{s}}_{\mathrm{JSCC}}, \gamma).
\end{equation}
Let $\mathbf{x}_0\sim\mathcal{N}(0,\mathbf{I})$ denote the Gaussian source sample and $\mathbf{x}_1=\mathbf{s}$ denote the clean target image. During training, we consider the linear conditional probability path as follows~\cite{liu2023rectifiedflow}:
\begin{equation}
\label{interpolation}
    \mathbf{x}_t=(1-t)\,\mathbf{x}_0+t\,\mathbf{x}_1,
\end{equation}
where $t\sim\mathcal{U}(0,1)$. Hence, $\mathbf{x}_t$ is pure noise at $t=0$ and yields the target image at $t=1$.



We train the velocity network in the proposed FlowSem using the following loss:
\begin{equation}
\label{fm_loss}
    \mathcal{L}_{\mathrm{FM}}
    =
    \mathbb{E}_{\mathbf{x}_0,\mathbf{x}_1,t}
    \left[
    \left\|
    v_{\psi}(\mathbf{x}_t,t,\hat{\mathbf{s}}_{\mathrm{JSCC}},\gamma)
    -(\mathbf{x}_1-\mathbf{x}_0)
    \right\|_2^2
    \right],
\end{equation}
where $v_{\psi}$ denotes the learned velocity field, conditioned on both $\hat{\mathbf{s}}_{\mathrm{JSCC}}$ and $\gamma$, and $\mathbf{x}_1-\mathbf{x}_0$ is the conditional target velocity. The conditioning information encourages the learned velocity to generate an image that is consistent with the coarse reconstruction recovered through the wireless channel and with the corresponding channel condition. As detailed in Fig.~\ref{fig:system_model}, we model the velocity field using a U-Net architecture~\cite{ronneberger2015unet}.



During inference, the ODE in~\eqref{ode_equation} is initialized using $mathbf{x}_0\sim\mathcal{N}(0,\mathbf{I})$ and numerically integrated from $t=0$ to $t = 1$. In FlowSem, we use the Euler method with $N_{\mathrm{step}}$ integration steps and step size $\Delta t=1/N_{\mathrm{step}}$ as follows:
\begin{equation}
    \mathbf{x}(t+\Delta t)
    =
    \mathbf{x}(t)
    +
    \Delta t\,
    v_{\psi}(\mathbf{x}(t),t,\hat{\mathbf{s}}_{\mathrm{JSCC}},\gamma).
\end{equation}

The proposed FlowSem demonstrates several benefits. First, within each considered channel model, FlowSem adapts to a wide range of SNR conditions using a single SNR-adaptive model rather than training a separate model for each SNR value. Second, it enhances the quality of the DeepJSCC reconstruction. Finally, it is evaluated using only a small number of ODE steps, thereby reducing the number of sampling steps compared with standard diffusion models. Empirically, we examine the quality-complexity tradeoff in Section~\ref{subsec:steps}. Algorithm~\ref{alg:overall} summarizes the proposed FlowSem algorithm.

\section{Experimental Results}
\label{sec:results}

This section presents the experimental results. First, we provide an overview on the dataset, channel settings, evaluation metrics and baselines. Then, we report numerical results through diverse simulations.

\subsection{Dataset and Communication Settings}

We consider the Cityscapes dataset~\cite{cordts2016cityscapes} to evaluate the proposed framework. The Cityscapes dataset contains high-resolution urban street-scene images with complex semantic and structural content such as roads, buildings, vehicles, pedestrians, and traffic signs. These characteristics make Cityscapes a challenging benchmark for reconstruction under compression and noisy channels.


We preprocess the images by resizing them into a fixed spatial resolution of $256 \times 256$ pixels, retaining all three RGB color channels, and normalizing the pixel intensities to [0, 1] to stabilize the training. In addition, we use the standard Cityscapes split comprising $2975$ training images and $500$ validation images. Two channel models are considered, AWGN and Rayleigh fading. For each considered channel model, FlowSem adapts to a wide range of SNR conditions using a single SNR-adaptive model rather than training a separate model for each SNR value. We consider an SNR range of $\{-10,-5,0,5,10,15,20,25\}$. The Cityscapes images contain $\mathrm{n}=3\times256\times256=196,608$ source values. We fix the same channel-symbol budget, \emph{i.e.}, fixed compression ratio $\mathrm{CR}=n/k=24$, across all methods to ensure fair comparison. This corresponds to a latent size of $\mathrm{k}=8192$ channel symbols per image.

We set the DeepJSCC batch size to $64$, while using a batch size of $16$ for the flow model training. For the learning rates, we use $1 \times 10^{-4}$ and $2 \times 10^{-4}$ rates for DeepJSCC and the flow model, respectively. We train the DeepJSCC model using $150$ epochs, while training the flow model for $100$ epochs. Both models are trained with the Adam optimizer~\cite{kingma2015adam}. All models are implemented in PyTorch and trained on an NVIDIA GeForce RTX 4080 GPU. All simulation settings are reported in Table~\ref{tab:hyperparams}.

\begin{table}[t!]
\centering
\caption{Experimental parameters and training hyperparameters.}
\label{tab:hyperparams}
\begin{tabular}{cc|cc}
\toprule
\textbf{Parameter} & \textbf{Value} & \textbf{Parameter} & \textbf{Value} \\
\midrule
\midrule
Train/val images & $2975/500$ & Image size & $256\times256\times3$\\
Pixel range & $[0,1]$ & Transmit power $P$ & $1$\\
Source values $n$ & $196608$ & Symbols $k$ & $8192$ \\
Compression ratio & $24$ & Euler steps & $5$ \\
DeepJSCC latent & $32\times16\times16$ & DeepJSCC epochs & $150$ \\
DeepJSCC batch & $64$ & DeepJSCC LR & $1\times10^{-4}$ \\
$\beta_t$ (Linear) & $10^{-4}-0.02$ & Weight $\delta$ & $0.5$ \\
Weight $\omega$ & $0.05$ & Diffusion LR & $2\times10^{-4}$ \\ 
Optimizer & Adam & Flow LR & $2\times10^{-4}$ \\
Flow batch size & $16$ & Flow epochs & $100$ \\
Diffusion epochs & $200$
& Diffusion batch size & $16$ \\
\bottomrule
\end{tabular}%
\end{table}

\subsection{Evaluation Metrics}\label{metrics}
The reconstruction methods are evaluated using distortion-based, structural, perceptual, and distribution-level metrics. Let $\mathbf{s}$ and $\hat{\mathbf{s}}$ denote the original and reconstructed images. The \textbf{mean squared error (MSE)} is calculated as follows:
\begin{equation}
    \mathrm{MSE}=\frac{1}{3HW}\left\|\mathbf{s}-\hat{\mathbf{s}}\right\|_2^2,
\end{equation}
while the \textbf{peak signal-to-noise ratio (PSNR)} is calculated as follows:
\begin{equation}
\mathrm{PSNR}=10\log_{10}\!\left(\frac{1}{\mathrm{MSE}}\right).
\end{equation}

Structural quality is measured using the \textbf{structural similarity index (SSIM)}~\cite{wang2004ssim}, which compares local luminance, contrast, and structure. It can be denoted as:
\begin{equation}
    \mathrm{SSIM}(x, y) = \frac{(2 \mu_{x} \mu_{y} + C_{1})(2 \sigma_{xy} + C_{2})}{(\mu_{x}^{2} + \mu_{y}^{2} + C_{1})(\sigma_{x}^{2} + \sigma_{y}^{2} + C_{2})}.
\end{equation}
Here $\mu_{x}, \mu_{y}$ are local means, $\sigma_{x}^{2}, \sigma_{y}^{2}$ are local variances, $\sigma_{xy}$ is the local covariance, and $C_{1}, C_{2}$ are small stability constants. Similarly, the \textbf{multi-scale SSIM (MS-SSIM)}~\cite{wang2003msssim} evaluates structural similarity across image scales.

Perceptual quality is evaluated using the \textbf{learned perceptual image patch similarity (LPIPS)}~\cite{zhang2018lpips}. LPIPS measures the perceptual distance between two images by comparing normalized deep feature representations extracted from a pretrained AlexNet model~\cite{krizhevsky2012alexnet}. The perceptual distance is computed as:
\begin{equation}
d_{\mathrm{LPIPS}}(\mathbf{x},\mathbf{y})
=
\sum_{l}
\frac{1}{H_lW_l}
\sum_{i,j}
\left\|
\mathbf{w}_l
\odot
\left(
\hat{\mathbf{x}}_{ij}^{\,l}
-
\hat{\mathbf{y}}_{ij}^{\,l}
\right)
\right\|_2^2,
\end{equation}
where $\hat{\mathbf{x}}_{ij}^{\,l}$ and $\hat{\mathbf{y}}_{ij}^{\,l}$ denote the normalized feature vectors extracted from layer $l$ at spatial location $(i,j)$, $\mathbf{w}_l$ represents the learned channel-wise weights, and $H_l$ and $W_l$ are the spatial dimensions of the feature map at layer $l$. Unlike pixel-wise metrics, LPIPS provides a learned measure of perceptual similarity between two images.

Distribution-level similarity is measured using the \textbf{Fr\'echet inception distance (FID)}~\cite{heusel2017fid}, which compares the statistics of generated and real images in the feature space of a pretrained Inception-V3 network as follows:
\begin{equation}
    \mathrm{FID}
    =
    \left\|\mu_r-\mu_g\right\|_2^2
    +
    \Tr\!\left(\Sigma_r+\Sigma_g-2(\Sigma_r\Sigma_g)^{1/2}\right),
\end{equation}
where $(\mu_r,\Sigma_r)$ and $(\mu_g,\Sigma_g)$ are the mean and covariance of the real-image and generated-image features, respectively. For clarity, lower MSE, LPIPS, and FID indicate better performance, whereas higher PSNR, SSIM, and MS-SSIM indicate better performance.

\subsection{Baselines}
The proposed conditional flow matching reconstruction is compared against four baselines as follows: 
\begin{enumerate}
    \item \textbf{DeepJSCC:} The first baseline is DeepJSCC, where we use the trained DeepJSCC as detailed in algorithm~\ref{alg:overall}. The output $\hat{\mathbf{s}}_{\mathrm{JSCC}}$ is used as the reconstructed image as described in Section~\ref{subsec:deepjscc}. This baseline serves as an ablation for evaluating the contribution of the second-stage (flow matching) in the reconstruction.

    \item \textbf{DiffSem-DDPM:} The second baseline is DDPM-based semantic communication (DiffSem-DDPM) model. In this baseline, we use the well-known DDPM~\cite{ho2020ddpm} model in the second stage instead of the proposed flow matching. DiffSem-DDPM uses the same pretrained DeepJSCC and the same U-Net backbone as the proposed FlowSem framework. This ensures a fair comparison between the generative approaches.

    \item \textbf{DiffSem-DDIM:} The third baseline is DDIM-based semantic communication (DiffSem-DDIM) model. In this baseline, we adopt the DDIM~\cite{song2021ddim} sampling model in the second stage. DiffSem-DDIM uses exactly the same trained DiffSem-DDPM, but replaces standard DDPM sampling with accelerated deterministic DDIM sampler. Unless otherwise stated, DiffSem-DDIM uses five sampling steps.

    \item \textbf{BPG+LDPC:} The fourth baseline is a conventional separation-based digital system that combines better portable graphics (BPG) image compression~\cite{bellard2018bpg} with low-density parity-check (LDPC) channel coding~\cite{gallager1962ldpc}. Each image is first compressed with BPG, and the resulting bitstream is LDPC-encoded and modulated using binary phase-shift keying (BPSK). 
    For a fair comparison, the baseline is rate-matched to the same channel-symbol budget $k$ used by the neural methods, so that the number of transmitted coded BPSK symbols does not exceed $k$. The BPG quality parameter is selected so that the compressed stream fits within the resulting payload budget. Under Rayleigh fading, perfect channel state information is assumed, zero-forcing equalization is applied, and symbol-wise log-likelihood ratios are passed to the LDPC decoder to account for the unequal reliability of faded symbols. If the compressed BPG bitstream exceeds the available payload budget or the LDPC decoder produces an invalid BPG bitstream, the transmission is declared unsuccessful and the reconstructed image is represented by a neutral grey image for metric computation.
\end{enumerate}


\begin{figure}[t!]
    \centering
    \subfloat[MSE]{\includegraphics[width=0.48\columnwidth,trim={0 0 0 0},clip]{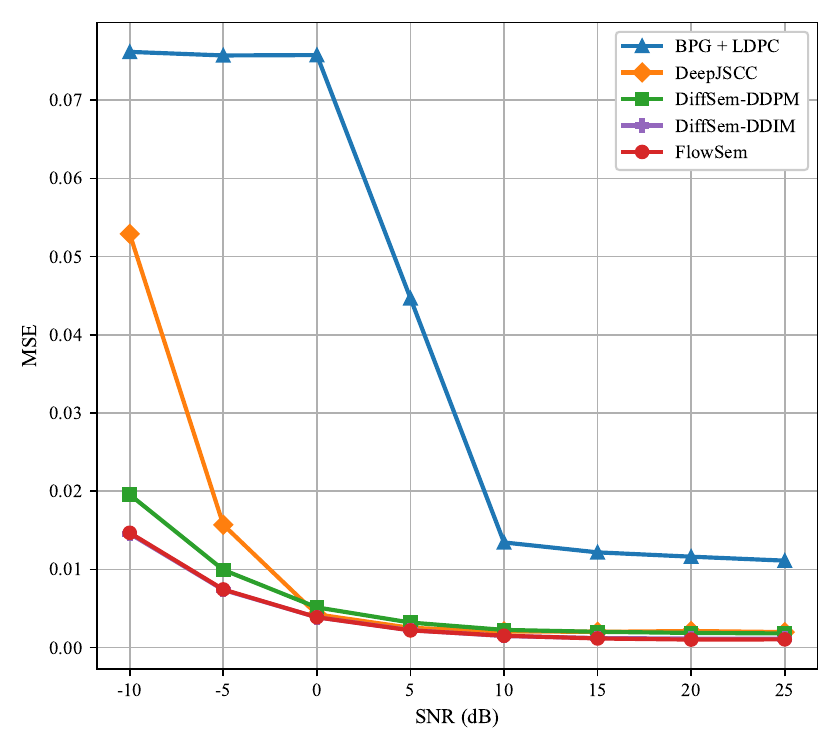}\label{mse_awgn}}
    \vspace{-1mm}
    \subfloat[PSNR (dB)]{\includegraphics[width=0.48\columnwidth,trim={0 0 0 0},clip]{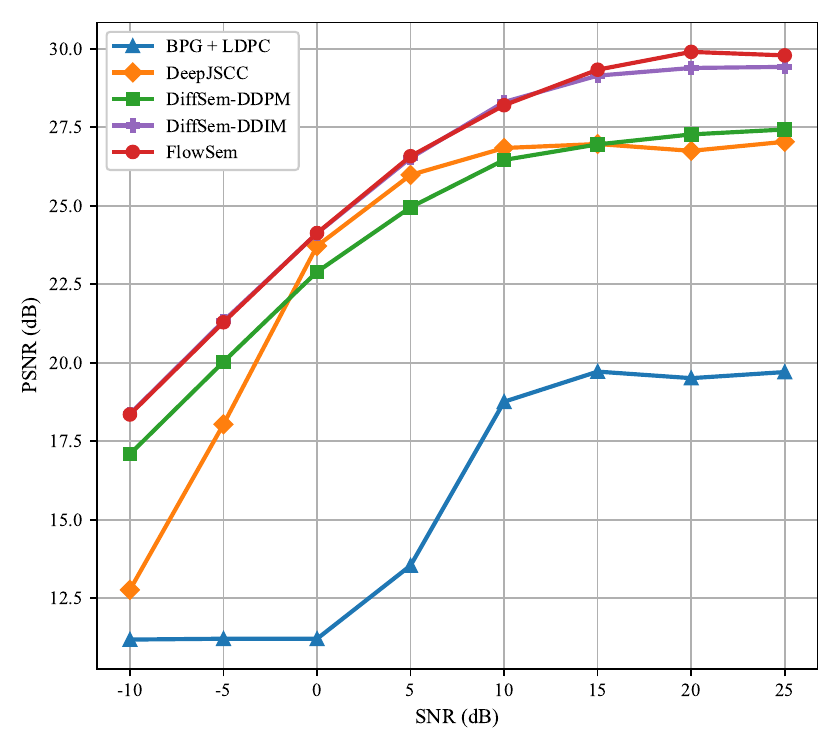}\label{psnr_awgn}} 
    \hfill
    \subfloat[SSIM]{\includegraphics[width=0.48\columnwidth,trim={0 0 0 0},clip]{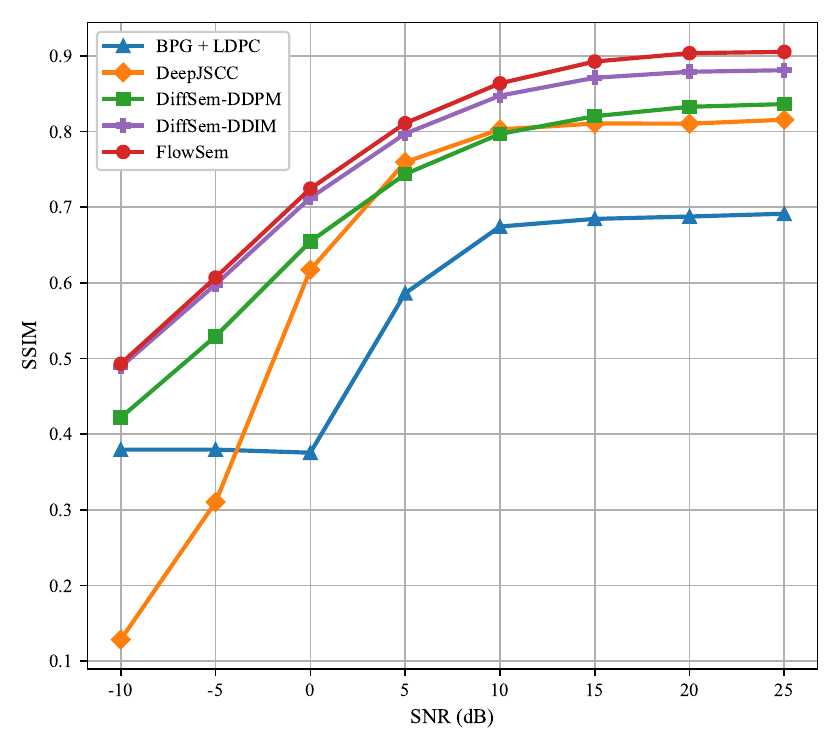}\label{ssim_awgn}}
    \vspace{-1mm}
    \subfloat[MS-SSIM]{\includegraphics[width=0.48\columnwidth,trim={0 0 0 0},clip]{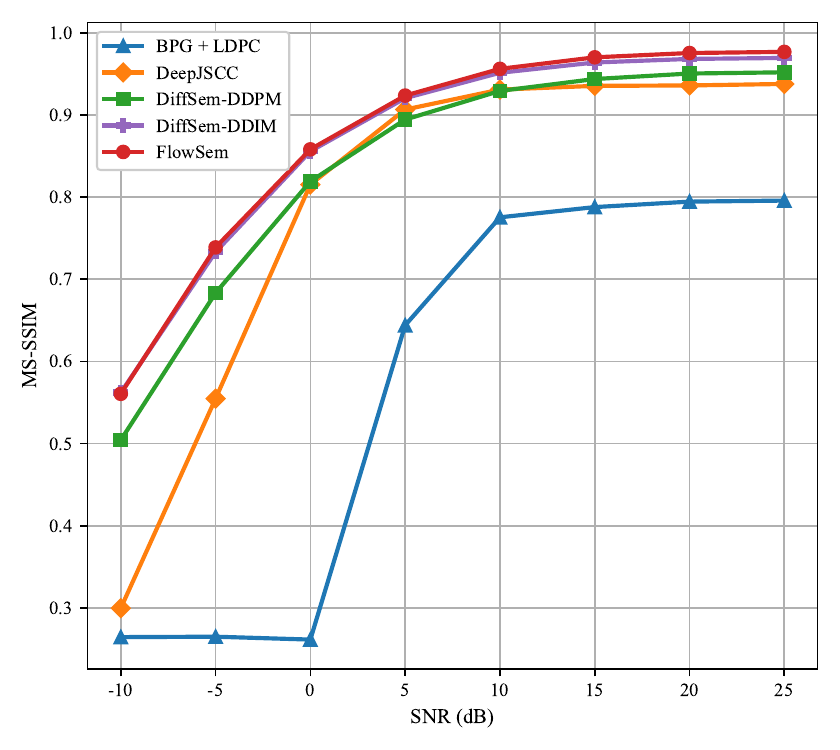}\label{msssim_awgn}}
    \hfill
    \subfloat[LPIPS]{\includegraphics[width=0.48\columnwidth,trim={0 0 0 0},clip]{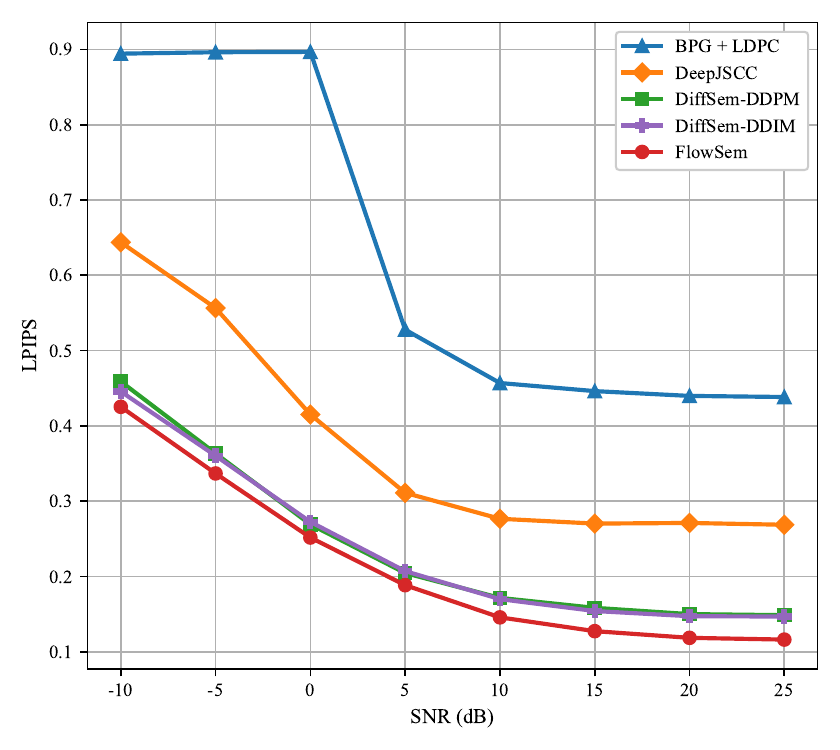}\label{lpips_awgn}}
    \vspace{-1mm}
    \subfloat[FID]{\includegraphics[width=0.48\columnwidth,trim={0 0 0 0},clip]{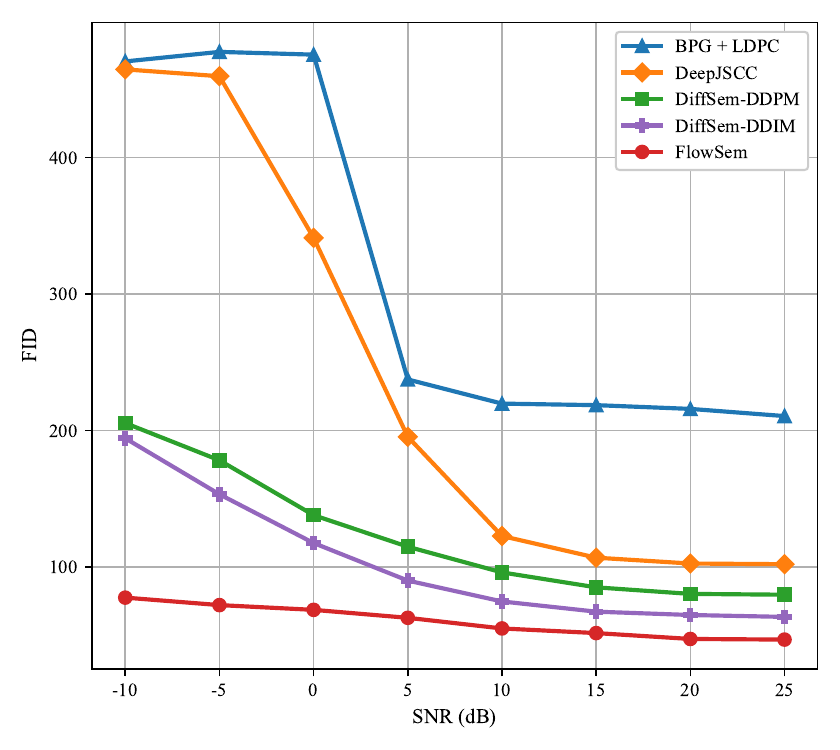}\label{fid_awgn}}
    \hfill
    \vspace{2mm}
    \caption{Reconstruction performance of FlowSem and the considered baselines as function of SNR under the AWGN channel.}
    \vspace{0mm}
    \label{fig:awgn_results}
\end{figure}

\begin{figure}[t!]
    \centering
    \subfloat[MSE]{\includegraphics[width=0.48\columnwidth,trim={0 0 0 0},clip]{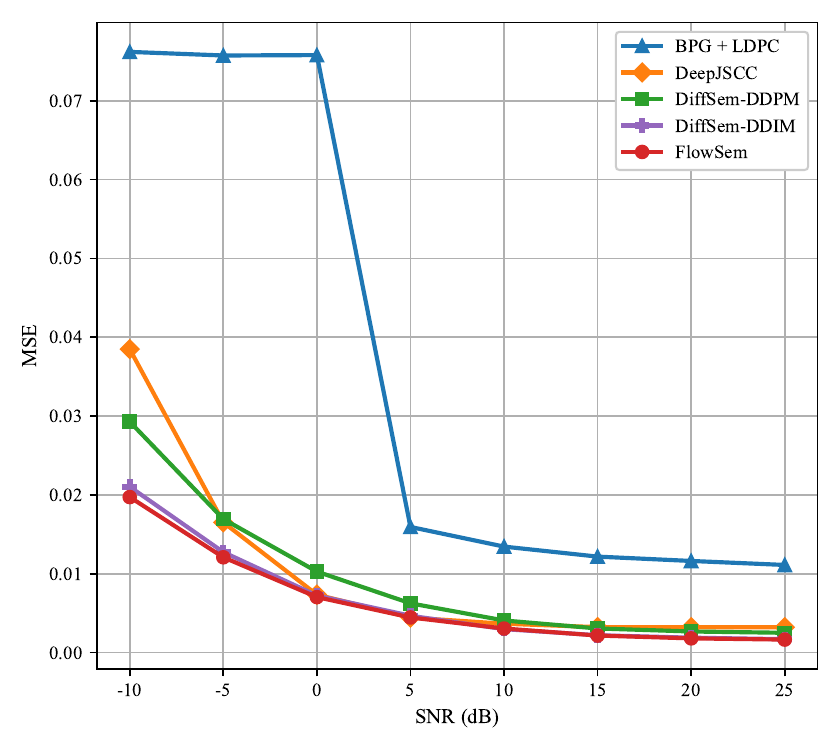}\label{mse_fading}}
    \vspace{-1mm}
    \subfloat[PSNR (dB)]{\includegraphics[width=0.48\columnwidth,trim={0 0 0 0},clip]{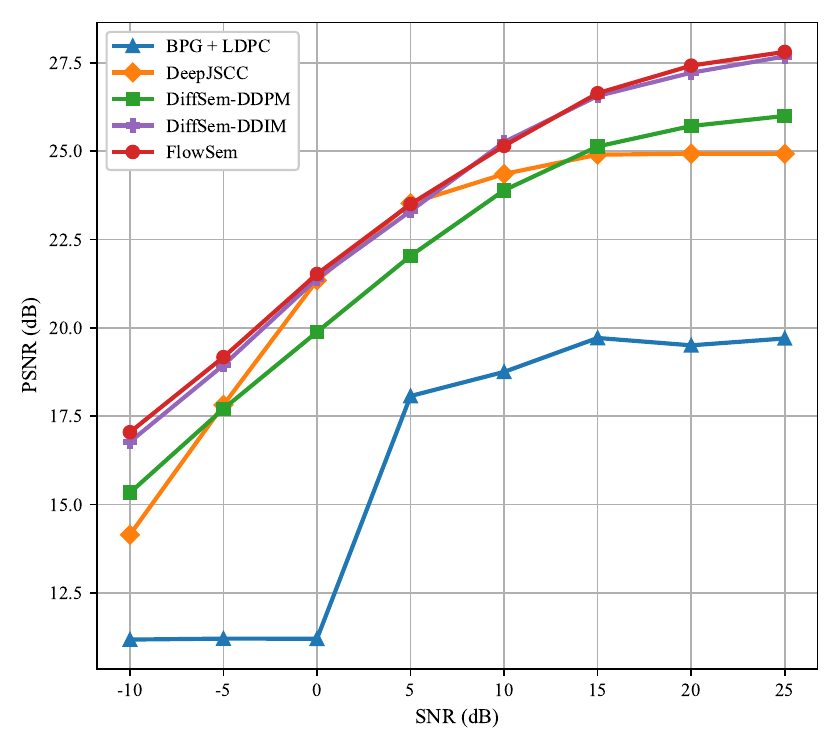}\label{psnr_fading}} 
    \hfill
    \subfloat[SSIM]{\includegraphics[width=0.48\columnwidth,trim={0 0 0 0},clip]{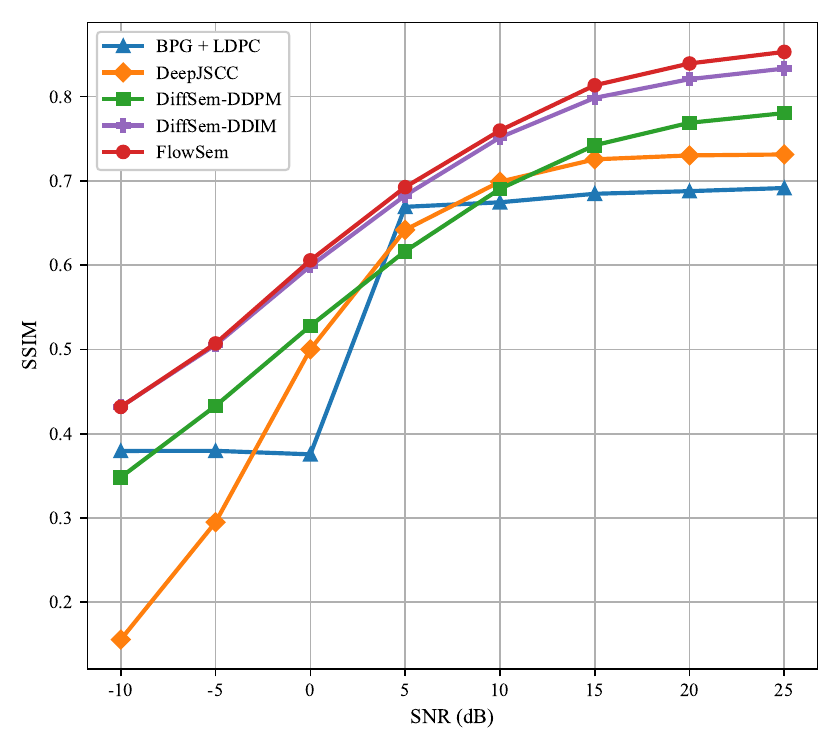}\label{ssim_fading}} 
    \vspace{-1mm}
    \subfloat[MS-SSIM]{\includegraphics[width=0.48\columnwidth,trim={0 0 0 0},clip]{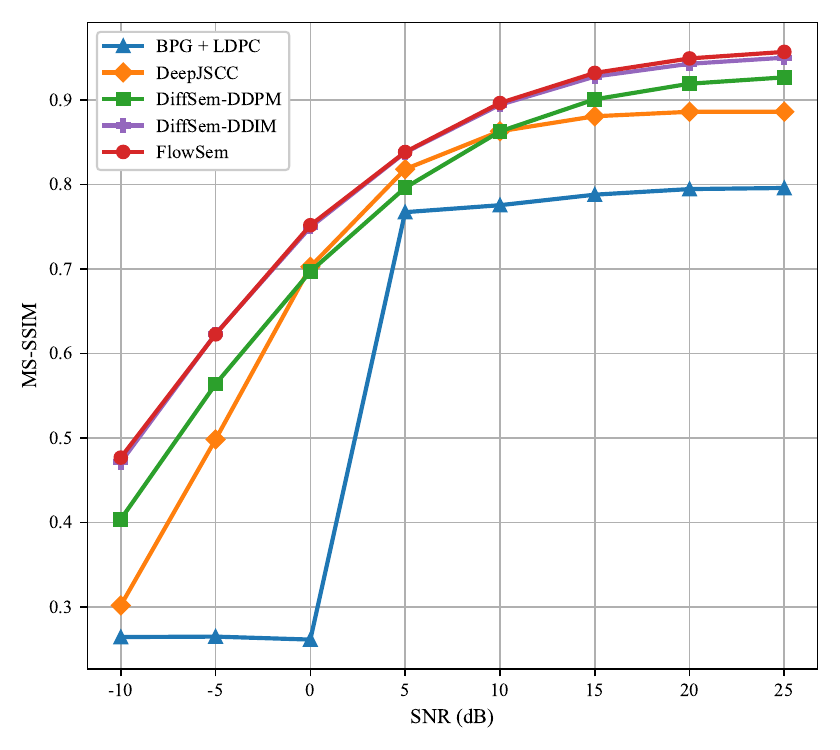}\label{msssim_fading}}
    \hfill
    \subfloat[LPIPS]{\includegraphics[width=0.48\columnwidth,trim={0 0 0 0},clip]{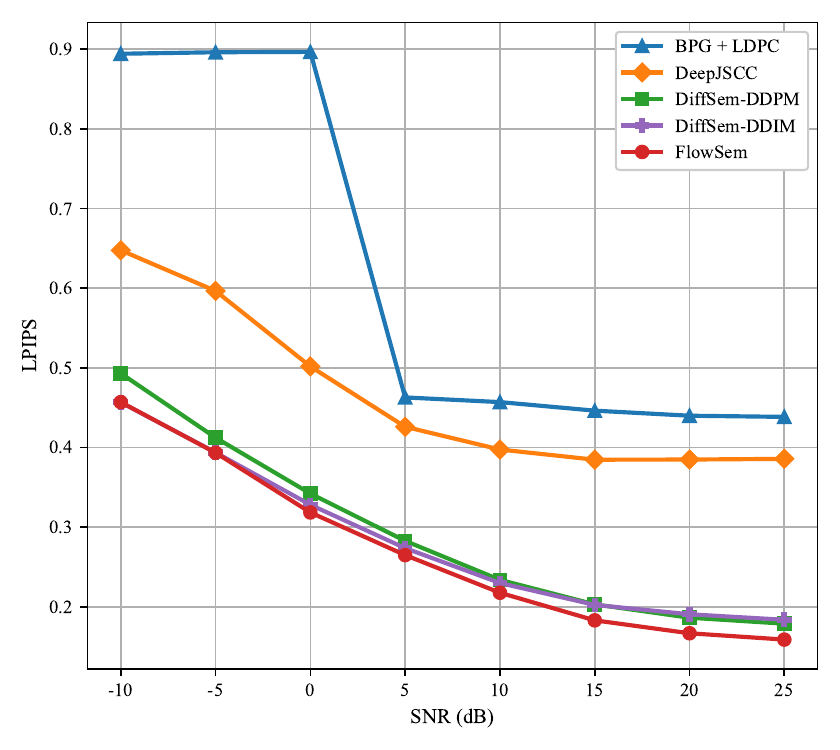}\label{lpips_fading}}
    \subfloat[FID]{\includegraphics[width=0.48\columnwidth,trim={0 0 0 0},clip]{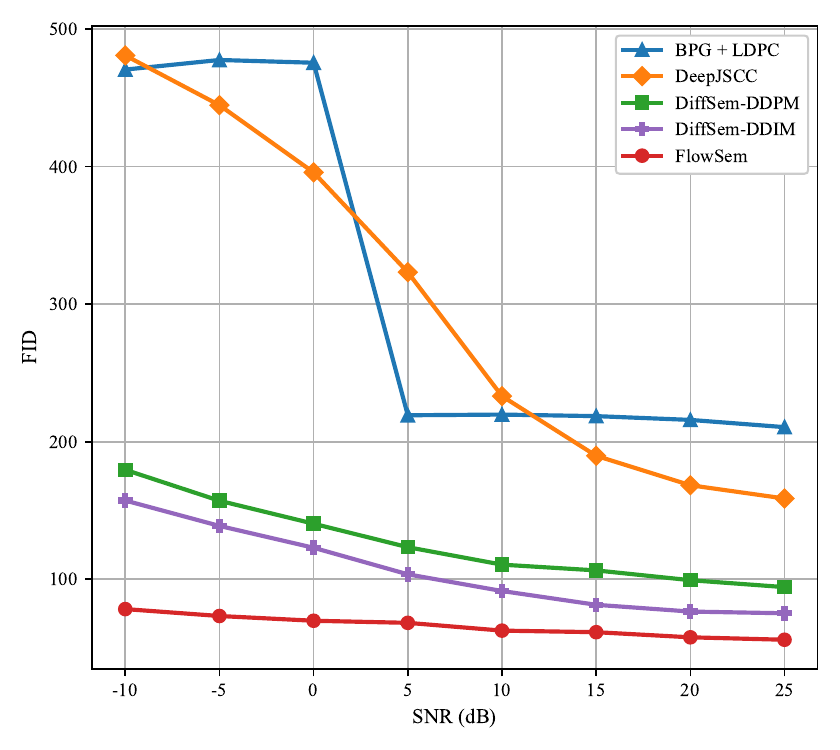}\label{fid_fading}}
    \hfill
    \vspace{2mm}
    \caption{Reconstruction performance of FlowSem and the considered baselines as function of SNR under the Rayleigh fading channel.}
    \vspace{0mm}
    \label{fig:fading_results}
\end{figure}

\subsection{Results on AWGN Channels}
In the first set of experiments in Fig.~\ref{fig:awgn_results}, we compare the proposed FlowSem to the baselines in terms of the metrics introduced in Section~\ref{metrics} under the AWGN channel. Figs.~\ref{mse_awgn} and~\ref{psnr_awgn} present the MSE and PSNR results, respectively. BPG+LDPC baseline exhibits the expected cliff-effect behavior at low SNRs, where decoding failures lead to a sharp degradation in reconstruction accuracy. DeepJSCC avoids this abrupt failure and shows relatively lower MSE and higher PSNR compared with BPG+LDPC. The diffusion baselines, \emph{i.e.}, DiffSem-DDPM and DiffSem-DDIM, further improve the reconstructed images compared with DeepJSCC over the considered SNR range. The proposed FlowSem achieves the lowest MSE and highest PSNR compared with all baselines over the considered SNR range.


Figs.~\ref{ssim_awgn} and~\ref{msssim_awgn} report the structural similarity metrics (SSIM and MS-SSIM). While DeepJSCC preserves the general structure of the source images, it loses part of the fine structural information especially at low SNRs. Both diffusion baselines improve the structural quality of the reconstructed images. The proposed FlowSem consistently achieves higher SSIM and MS-SSIM than the considered baselines. The gap in structural similarity compared with DiffSem-DDIM is more noticeable in high SNR regions. This highlights better structural similarities of the reconstructed images compared with the baselines.


Finally, Figs.~\ref{lpips_awgn} and~\ref{fid_awgn} report LPIPS and FID metrics. The proposed FlowSem demonstrates the largest improvement observed in these two metrics. At $-10$ dB SNR, FlowSem reduces FID by approximately $60 \%$ relative to the best-performing diffusion baseline. This highlights the benefits of introducing flow matching as a refinement stage after DeepJSCC. In addition, it shows the advantage of FlowSem in improving perceptual similarity and distribution-level image quality compared with the baselines.

\subsection{Results on Rayleigh Fading Channels}
In the next experiment in Fig.~\ref{fig:fading_results}, we examine the performance under Rayleigh fading channels. Rayleigh fading is more challenging than AWGN, because each transmitted symbol is affected by a random fading coefficient in addition to the additive noise. As a result, symbols experience different instantaneous channel conditions even when the average SNR is the same. Similar to the AWGN experiment, BPG+LDPC exhibits the cliff-effect at low SNRs. This leads to large performance degradation across all reported metrics.

In Figs.~\ref{mse_fading} and~\ref{psnr_fading}, we observe that DeepJSCC exhibits larger degradation at low SNRs than the generative baselines. DiffSem-DDPM improves the MSE and PSNR relative to DeepJSCC but remains below FlowSem. DiffSem-DDIM and FlowSem achieve similar pixel-wise reconstruction performance, with FlowSem showing a small advantage particularly at low SNRs.


The structural-quality metrics are reported in Figs.~\ref{ssim_fading} and~\ref{msssim_fading}. The proposed FlowSem consistently achieves higher SSIM and MS-SSIM than DeepJSCC, DiffSem-DDPM and DiffSem-DDIM. This indicates improved preservation of the image structure under fading. Similar to the AWGN case, FlowSem demonstrates the largest performance gap in terms of SSIM at high SNR regions.


\begin{figure*}[t!]
    \centering
    \includegraphics[width=1.6\columnwidth,trim={0 0 0 0},clip]{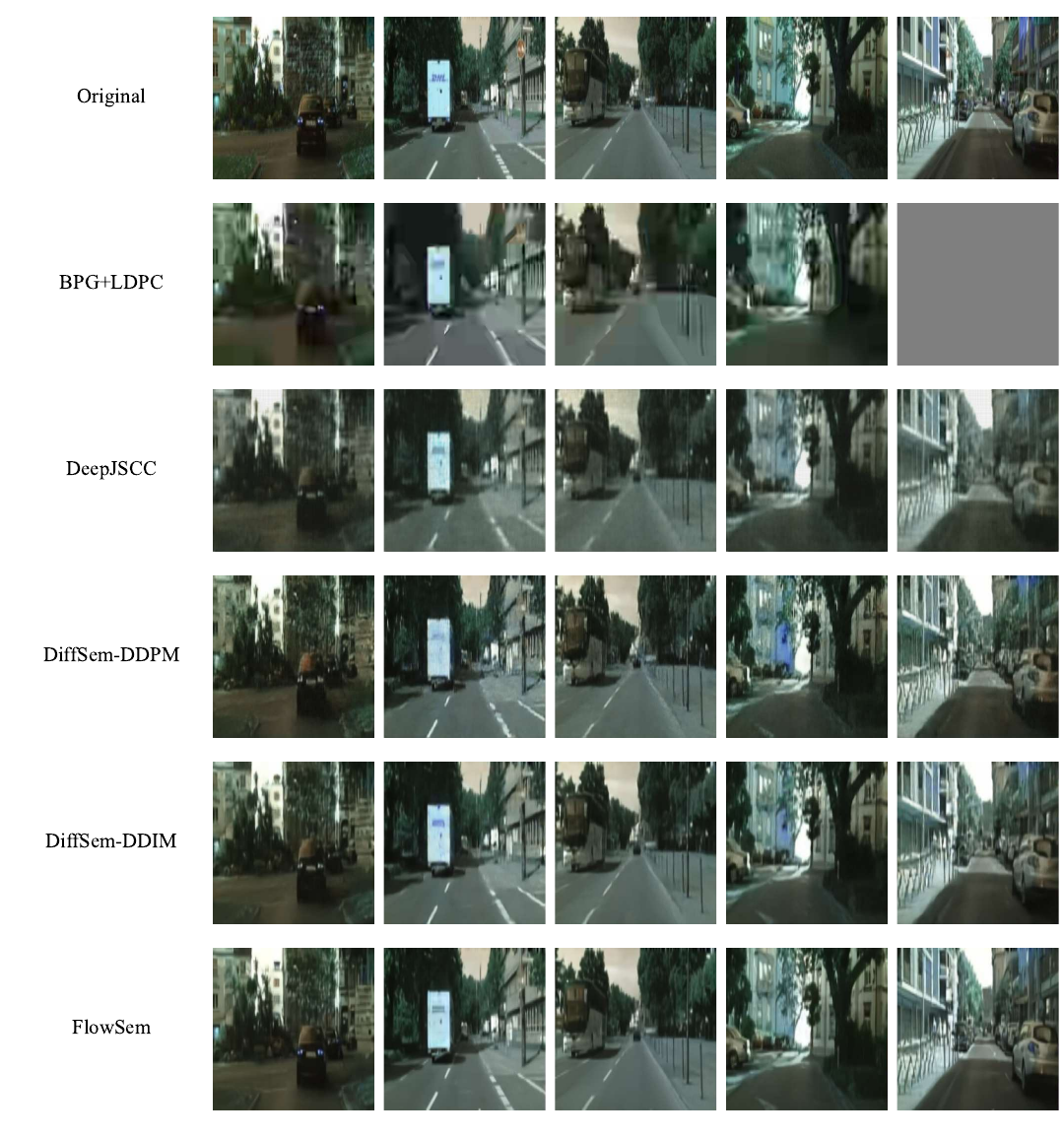}
    \\[-1mm]
    \caption{Representative reconstructed images under the AWGN channel at an SNR of $10$ dB.}
    \vspace{0mm}
    \label{fig:visual_awgn}
\end{figure*}

We show the perceptual and distribution-level metrics in Figs.~\ref{lpips_fading} and ~\ref{fid_fading}. The performance gap between DeepJSCC and the generative methods becomes more pronounced under severe fading conditions. The proposed FlowSem achieves the lowest LPIPS and FID over the considered SNR range. At $-10$ dB SNR, FlowSem reduces FID by approximately $40 \%$ relative to the best-performing diffusion baseline. This demonstrates that the flow matching refinement remains effective when the coarse DeepJSCC reconstruction is affected by fading.



\subsection{Visual Comparison}




In the next experiment, we present representative examples of reconstructed images under the AWGN channel at an SNR of $10$ dB in Fig.~\ref{fig:visual_awgn}. We can observe that BPG + LDPC generally produces lower-quality reconstruction under the imposed channel-symbol budget. While DeepJSCC successfully recovers the main scene structure of the original images, it loses some fine visual details. Both diffusion baselines (DiffSem-DDPM and DiffSem-DDIM) provide sharper reconstruction compared with DeepJSCC. However, they still exhibit visible reconstruction artifacts in several image regions. In contrast, the proposed FlowSem provides the clearest visual reconstruction among the considered methods in these examples. In particular, it preserves important structures including road boundaries, vehicles, and background objects. These qualitative observations are consistent with the improvements observed in LPIPS and FID metrics.

\begin{table*}[t]
\centering
\caption{Reconstruction quality as a function of the number of sampling steps at
$\mathrm{CR}=24$ and SNR $=10$ dB under AWGN and fading channels.}
\label{tab:steps_channels}
\setlength{\tabcolsep}{3.1pt}
\renewcommand{\arraystretch}{1.15}
\begin{tabular}{c|c|cccccc|cccccc}
\toprule
\multirow{2}{*}{\textbf{Model}} &
\multirow{2}{*}{\textbf{Steps}} &
\multicolumn{6}{c|}{\textbf{AWGN}} &
\multicolumn{6}{c}{\textbf{Fading}} \\
\cmidrule(lr){3-8}\cmidrule(lr){9-14}
& &
MSE$\downarrow$ &
PSNR$\uparrow$ &
SSIM$\uparrow$ &
MS-SSIM$\uparrow$ &
LPIPS$\downarrow$ &
FID$\downarrow$ &
MSE$\downarrow$ &
PSNR$\uparrow$ &
SSIM$\uparrow$ &
MS-SSIM$\uparrow$ &
LPIPS$\downarrow$ &
FID$\downarrow$ \\
\midrule

\multirow{4}{*}{DiffSem-DDPM}
& $100$
& $0.19597$ & $6.915$ & $0.0058$ & $0.0575$ & $1.5771$ & $523.608$
& $0.19591$ & $7.080$ & $0.0059$ & $0.0573$ & $1.5813$ & $521.469$ \\

& $500$
& $0.16816$ & $7.596$ & $0.0163$ & $0.1644$ & $1.5670$ & $491.535$
& $0.16840$ & $7.737$ & $0.0160$ & $0.1600$ & $1.5722$ & $493.947$ \\

& $750$
& $0.09434$ & $10.191$ & $0.0527$ & $0.3862$ & $1.4884$ & $414.726$
& $0.09517$ & $10.215$ & $0.0498$ & $0.3694$ & $1.4915$ & $412.541$ \\

& $1000$
& $0.00258$ & $26.635$ & $0.7883$ & $0.9289$ & $0.1475$ & $117.310$
& $0.00389$ & $24.108$ & $0.7154$ & $0.8826$ & $0.2229$ & $135.148$ \\

\midrule

\multirow{6}{*}{DiffSem-DDIM}
& $5$
& $0.00152$ & $28.195$ & $0.8476$ & $0.9513$ & $0.1695$ & $75.107$
& $\mathbf{0.00252}$ & $\mathbf{25.998}$ & $0.7762$ & $0.9129$ & $0.2267$ & $89.347$ \\

& $10$
& $0.00172$ & $27.637$ & $0.8329$ & $0.9451$ & $0.1618$ & $80.544$
& $0.00285$ & $25.475$ & $0.7579$ & $0.9034$ & $0.2138$ & $88.359$ \\

& $25$
& $0.00194$ & $27.124$ & $0.8140$ & $0.9378$ & $0.1563$ & $92.936$
& $0.00315$ & $25.029$ & $0.7364$ & $0.8929$ & $0.2061$ & $99.772$ \\

& $50$
& $0.00204$ & $26.906$ & $0.8035$ & $0.9339$ & $0.1596$ & $100.471$
& $0.00331$ & $24.823$ & $0.7237$ & $0.8873$ & $0.2082$ & $110.459$ \\

& $75$
& $0.00208$ & $26.813$ & $0.7993$ & $0.9324$ & $0.1624$ & $102.862$
& $0.00335$ & $24.759$ & $0.7196$ & $0.8855$ & $0.2111$ & $113.107$ \\

& $100$
& $0.00211$ & $26.768$ & $0.7971$ & $0.9316$ & $0.1646$ & $103.245$
& $0.00339$ & $24.713$ & $0.7173$ & $0.8846$ & $0.2124$ & $113.902$ \\

\midrule

\multirow{6}{*}{FlowSem}
& $5$
& $\mathbf{0.00145}$ & $\mathbf{28.398}$ & $\mathbf{0.8665}$
& $\mathbf{0.9567}$ & $0.1453$ & $\mathbf{56.483}$
& $0.00253$ & $25.981$ & $\mathbf{0.7899}$
& $\mathbf{0.9169}$ & $0.2058$ & $\mathbf{64.961}$ \\

& $10$
& $0.00180$ & $27.893$ & $0.8509$ & $0.9496$ & $0.1400$ & $68.727$
& $0.00313$ & $25.049$ & $0.7645$ & $0.9053$ & $0.1945$ & $77.879$ \\

& $25$
& $0.00202$ & $27.377$ & $0.8323$ & $0.9423$ & $0.1343$ & $62.205$
& $0.00347$ & $24.602$ & $0.7424$ & $0.8948$
& $\mathbf{0.1897}$ & $71.531$ \\

& $50$
& $0.00213$ & $27.135$ & $0.8219$ & $0.9383$
& $\mathbf{0.1339}$ & $61.445$
& $0.00363$ & $24.408$ & $0.7308$ & $0.8894$
& $\mathbf{0.1897}$ & $70.780$ \\

& $75$
& $0.00217$ & $27.046$ & $0.8173$ & $0.9366$ & $0.1343$ & $61.763$
& $0.00369$ & $24.336$ & $0.7261$ & $0.8873$ & $0.1905$ & $71.207$ \\

& $100$
& $0.00219$ & $26.999$ & $0.8151$ & $0.9358$ & $0.1348$ & $62.066$
& $0.00372$ & $24.301$ & $0.7237$ & $0.8863$ & $0.1910$ & $71.654$ \\

\bottomrule
\end{tabular}
\end{table*}

\begin{figure*}[!t]
    \centering
    \subfloat[DiffSem-DDPM]{%
        \includegraphics[width=1.9\columnwidth,trim={0 0 0 0},clip]{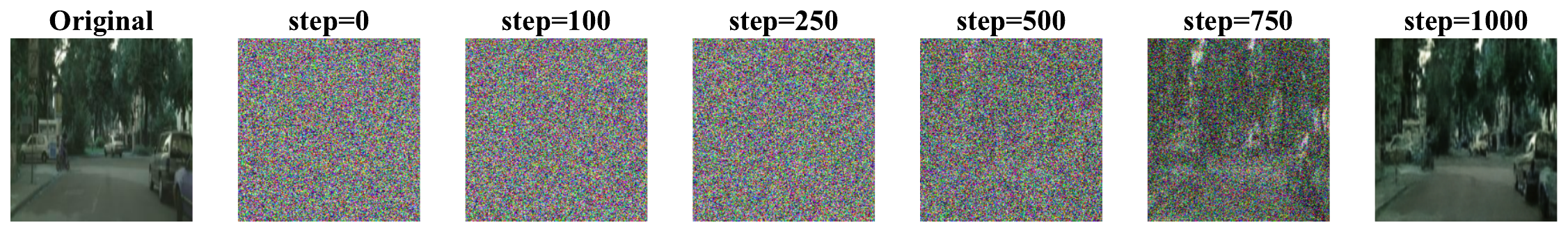}
        \label{fig:DiffSem}
    }
    \\[1mm]
    \subfloat[DiffSem-DDIM]{%
        \includegraphics[width=1.9\columnwidth,trim={0 0 0 0},clip]{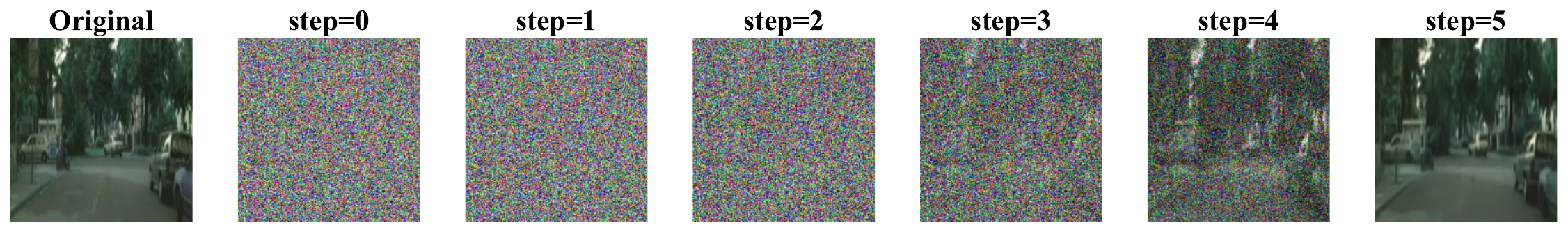}
        \label{fig:DiffSem-DDIM}
    }
    \\[1mm]
    \subfloat[FlowSem]{%
        \includegraphics[width=1.9\columnwidth,trim={0 0 0 0},clip]{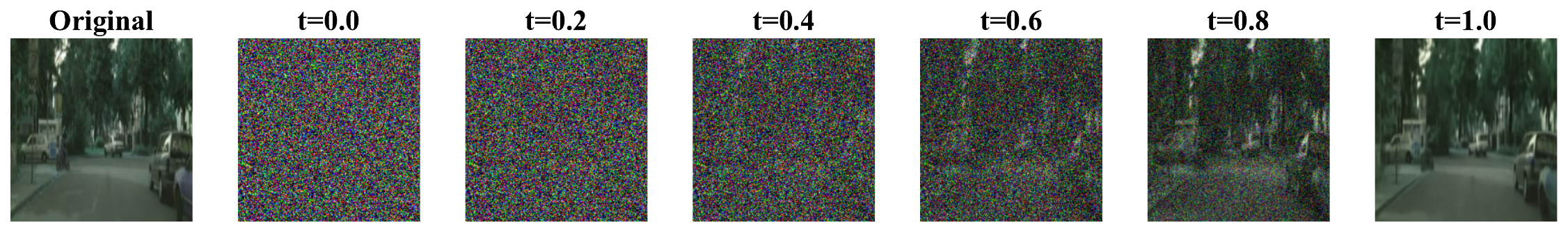}
        \label{fig:FlowSem-DDM}
    }
    \vspace{1mm}
    \caption{Evolution of the reconstructed image during DiffSem-DDPM, DiffSem-DDIM, and FlowSem sampling.}
    \vspace{0mm}
    \label{fig:steps}
\end{figure*}

\subsection{Number of Sampling Steps Effect}
\label{subsec:steps}
 
The main motivation for adopting flow matching is its sampling efficiency. To investigate this property, we evaluate FlowSem compared with DiffSem-DDPM and DiffSem-DDIM as a function of the number of generative-network evaluations in Table~\ref{tab:steps_channels} under AWGN and fading channels at an SNR of $10$ dB. DiffSem-DDPM requires around $1000$ sampling steps to provide good pixel-wise reconstruction while maintaining improved perception performance as observed in the reported metrics. DiffSem-DDIM substantially reduces the number of network evaluations required compared with DiffSem-DDPM. While DiffSem-DDIM achieves strong pixel-wise reconstruction performance, its LPIPS and FID values remain consistently higher than those of FlowSem. This indicates that the main advantage of FlowSem at matched network evaluations lies in its perceptual and distribution-level reconstruction quality.

We observe that FlowSem reaches strong reconstruction quality using only five Euler steps. Increasing the number of Euler steps beyond five introduces a fidelity-perception tradeoff. In particular, pixel-wise and structural fidelity gradually decrease, as indicated by increasing MSE and decreasing PSNR, SSIM, and MS-SSIM. In contrast, LPIPS initially improves before saturating around $50$ steps. We set the number of Euler steps to $5$ as the default FlowSem configuration as it provides the best tradeoff between reconstruction fidelity and inference complexity in the considered experiments.

The proposed FlowSem achieves the best MSE, PSNR, SSIM, MS-SSIM, and FID using $5$ sampling steps, while achieving the best LPIPS at $50$ sampling steps. The best results are highlighted in bold in Table~\ref{tab:steps_channels}. At the same number of network evaluations, FlowSem consistently outperforms DiffSem-DDIM under AWGN. Under Rayleigh fading, both methods achieve comparable pixel-wise fidelity, while FlowSem provides higher structural similarity and substantially better perceptual and distribution-level quality. Fig.~\ref{fig:steps} shows the evolution of the reconstructed image from noise to the target image. We can observe that DiffSem-DDPM requires more than $750$ steps to show image-related details and $1000$ steps to recover the image. In contrast, both DiffSem-DDIM and FlowSem require only five steps to recover the image. These results show that the main advantage of FlowSem should be interpreted in terms of its quality–latency tradeoff relative to DiffSem-DDIM, rather than only its speedup relative to DiffSem-DDPM.

\subsection{Complexity Analysis}
\label{subsec:inference_time}

In this experiment, we present the complexity of the proposed FlowSem compared with the baselines. FlowSem, DiffSem-DDIM, and DiffSem-DDPM use the same U-Net backbone and therefore have comparable computational cost per generative-network evaluation. Their total inference complexity, however, depends strongly on the number of sampling steps. Standard DiffSem-DDPM incurs substantially larger inference latency because it uses $1000$ network evaluations. DiffSem-DDIM reuses the same trained diffusion model but reduces the inference cost through accelerated five-step sampling. FlowSem also uses five network evaluations and therefore has a complexity profile comparable to DiffSem-DDIM. While DiffSem-DDIM and FlowSem have similar complexity profiles, FlowSem achieves comparable pixel-level fidelity while providing improved structural, perceptual, and distribution-level reconstruction quality.



\begin{table*}[!t]
\centering
\caption{Complexity analysis of the proposed FlowSem compared with the baselines.}
\label{tab:inference_time}
\begin{tabular}{@{}lccccc@{}}
\toprule
\textbf{Metric} & \textbf{BPG+LDPC} & \textbf{DeepJSCC} & \textbf{DiffSem-DDPM} & \textbf{DiffSem-DDIM} & \textbf{FlowSem} \\
\midrule
Sampling steps & -- & -- & $1000$ & $5$ & $5$ \\
Inference time (s) & $0.4486$ & $0.003962$ & $19.10$ & $0.10$ & $0.10$ \\
Training time (h) & -- & $5.20$ & $8.32$ & $8.32$ & $6.67$ \\
FLOPs / evaluation & -- & $8.389G$ & $117.99G$ & $117.99G$ & $117.99G$ \\

\bottomrule
\end{tabular}
\end{table*}

\section{Conclusion}
\label{sec:conclusion}
This paper proposed FlowSem, a two-stage generative semantic communication framework for wireless image transmission under a stringent channel-symbol budget. In the first stage, an SNR-adaptive DeepJSCC model transmits the image and produces a coarse reconstruction at the receiver. In the second stage, a conditional flow matching model generates the final reconstruction from Gaussian noise using both the DeepJSCC output and the channel SNR as conditioning information. FlowSem was evaluated on the Cityscapes dataset under AWGN and Rayleigh fading channels. We considered several baselines including rate-matched BPG+LDPC, DeepJSCC, and diffusion-based generative reconstruction, \emph{i.e.}, DiffSem-DDPM and DiffSem-DDIM.

The results show that the generative refinement stage improves reconstruction quality over the standalone DeepJSCC receiver, particularly in structural, perceptual, and distribution-level metrics. Among the considered generative approaches, FlowSem achieves a favorable reconstruction quality while requiring only a small number of ODE integration steps. The sampling-step analysis further shows that FlowSem provides a good tradeoff between pixel-level fidelity, perceptual quality, and receiver-side complexity. Compared with DiffSem-DDPM and DiffSem-DDIM, FlowSem achieves up to $60 \%$ lower FID in low SNR ranges. Overall, the results demonstrate the potential of flow matching as an efficient generative refinement mechanism for adaptive wireless image transmission. Future research will investigate more accelerated flow models using knowledge distillation or shortcut models to further enhance the sampling time while preserving the reconstruction performance.

\bibliographystyle{IEEEtran}
\bibliography{references}

@article{qin2021semantic,
  author  = {Qin, Zhijin and Tao, Xiaoming and Lu, Jianhua and Tong, Wen and Li, Geoffrey Ye},
  title   = {Semantic Communications: Principles and Challenges},
  journal = {arXiv preprint arXiv:2201.01389},
  year    = {2022}
}

@article{xie2021deepsc,
  author  = {Xie, Huiqiang and Qin, Zhijin and Li, Geoffrey Ye and Juang, Biing-Hwang},
  title   = {Deep Learning Enabled Semantic Communication Systems},
  journal = {IEEE Transactions on Signal Processing},
  volume  = {69},
  pages   = {2663--2675},
  year    = {2021},
  doi     = {10.1109/TSP.2021.3071210}
}

@article{bourtsoulatze2019deepjscc,
  author  = {Bourtsoulatze, Eirina and Burth Kurka, David and G\"und\"uz, Deniz},
  title   = {Deep Joint Source-Channel Coding for Wireless Image Transmission},
  journal = {IEEE Transactions on Cognitive Communications and Networking},
  volume  = {5},
  number  = {3},
  pages   = {567--579},
  year    = {2019},
  doi     = {10.1109/TCCN.2019.2919300}
}

@article{xu2022adjscc,
  author  = {Xu, Jialong and Ai, Bo and Chen, Wei and Yang, Ang and Sun, Peng and Rodrigues, Miguel},
  title   = {Wireless Image Transmission Using Deep Source Channel Coding With Attention Modules},
  journal = {IEEE Transactions on Circuits and Systems for Video Technology},
  volume  = {32},
  number  = {4},
  pages   = {2315--2328},
  year    = {2022},
  doi     = {10.1109/TCSVT.2021.3082521}
}

@inproceedings{ding2021snradaptive,
  author    = {Ding, Mingze and Li, Jiahui and Ma, Mengyao and Fan, Xiaopeng},
  title     = {{SNR}-Adaptive Deep Joint Source-Channel Coding for Wireless Image Transmission},
  booktitle = {ICASSP 2021 -- IEEE International Conference on Acoustics, Speech and Signal Processing},
  pages     = {1555--1559},
  year      = {2021},
  doi       = {10.1109/ICASSP39728.2021.9414037}
}

@article{erdemir2023generativejscc,
  author  = {Erdemir, Ecenaz and Tung, Tze-Yang and Dragotti, Pier Luigi and G\"und\"uz, Deniz},
  title   = {Generative Joint Source-Channel Coding for Semantic Image Transmission},
  journal = {IEEE Journal on Selected Areas in Communications},
  volume  = {41},
  number  = {8},
  pages   = {2645--2657},
  year    = {2023},
  doi     = {10.48550/arXiv.2211.13772}
}

@inproceedings{wang2022perceptual,
  author    = {Wang, Jun and Wang, Sixian and Dai, Jincheng and Si, Zhongwei and Zhou, Dekun and Niu, Kai},
  title     = {Perceptual Learned Source-Channel Coding for High-Fidelity Image Semantic Transmission},
  booktitle = {GLOBECOM 2022 -- IEEE Global Communications Conference},
  pages     = {3959--3964},
  year      = {2022},
  organization = {IEEE}
}

@inproceedings{kurka2019successive,
  author    = {Burth Kurka, David and G{\"u}nd{\"u}z, Deniz},
  title     = {Successive Refinement of Images with Deep Joint Source-Channel Coding},
  booktitle = {IEEE 20th International Workshop on Signal Processing Advances in Wireless Communications (SPAWC)},
  pages     = {1--5},
  year      = {2019},
  doi       = {10.1109/SPAWC.2019.8815416}
}

@article{kurka2020deepjsccf,
  author  = {Burth Kurka, David and G{\"u}nd{\"u}z, Deniz},
  title   = {{DeepJSCC-f}: Deep Joint Source-Channel Coding of Images with Feedback},
  journal = {IEEE Journal on Selected Areas in Information Theory},
  volume  = {1},
  number  = {1},
  pages   = {178--193},
  year    = {2020},
  doi     = {10.1109/JSAIT.2020.2987203}
}

@inproceedings{yang2021ofdm,
  author    = {Yang, Mingyu and Bian, Chenghong and Kim, Hun-Seok},
  title     = {Deep Joint Source Channel Coding for Wireless Image Transmission with {OFDM}},
  booktitle = {ICC 2021 -- IEEE International Conference on Communications},
  pages     = {1--6},
  year      = {2021},
  doi       = {10.1109/ICC42927.2021.9500996}
}

@inproceedings{yang2022adaptivejscc,
  author    = {Yang, Mingyu and Kim, Hun-Seok},
  title     = {Deep Joint Source-Channel Coding for Wireless Image Transmission with Adaptive Rate Control},
  booktitle = {ICASSP 2022 -- IEEE International Conference on Acoustics, Speech and Signal Processing},
  pages     = {5193--5197},
  year      = {2022},
  doi       = {10.1109/ICASSP43922.2022.9746335}
}

@inproceedings{bian2023deepjscclpp,
  author    = {Bian, Chenghong and Shao, Yulin and G{\"u}nd{\"u}z, Deniz},
  title     = {{DeepJSCC-l++}: Robust and Bandwidth-Adaptive Wireless Image Transmission},
  booktitle = {GLOBECOM 2023 -- IEEE Global Communications Conference},
  pages     = {3148--3154},
  year      = {2023},
  doi       = {10.1109/GLOBECOM54140.2023.10436878}
}

@article{wu2023deepjsccmimo,
  author  = {Wu, Haotian and Shao, Yulin and Bian, Chenghong and Mikolajczyk, Krystian and G{\"u}nd{\"u}z, Deniz},
  title   = {Deep Joint Source-Channel Coding for Adaptive Image Transmission over {MIMO} Channels},
  journal = {IEEE Transactions on Wireless Communications},
  volume  = {23},
  number  = {10},
  pages   = {15002--15017},
  year    = {2024},
  doi     = {10.1109/TWC.2024.3422794}
}

@article{gunduz2023beyond,
  author  = {G\"und\"uz, Deniz and Qin, Zhijin and Aguerri, Inaki Estella and Dhillon, Harpreet S. and Yang, Zhaohui and Yener, Aylin and Wong, Kai Kit and Chae, Chan-Byoung},
  title   = {Beyond Transmitting Bits: Context, Semantics, and Task-Oriented Communications},
  journal = {IEEE Journal on Selected Areas in Communications},
  volume  = {41},
  number  = {1},
  pages   = {5--41},
  year    = {2023},
  doi     = {10.1109/JSAC.2022.3223408}
}

@article{dai2022ntscc,
  author  = {Dai, Jincheng and Wang, Sixian and Tan, Kailin and Si, Zhongwei and Qin, Xiaoqi and Niu, Kai and Zhang, Ping},
  title   = {Nonlinear Transform Source-Channel Coding for Semantic Communications},
  journal = {IEEE Journal on Selected Areas in Communications},
  volume  = {40},
  number  = {8},
  pages   = {2300--2316},
  year    = {2022},
  doi     = {10.1109/JSAC.2022.3180802}
}

@inproceedings{ho2020ddpm,
  author    = {Ho, Jonathan and Jain, Ajay and Abbeel, Pieter},
  title     = {Denoising Diffusion Probabilistic Models},
  booktitle = {Advances in Neural Information Processing Systems (NeurIPS)},
  volume    = {33},
  pages     = {6840--6851},
  year      = {2020}
}

@inproceedings{song2021ddim,
  author    = {Song, Jiaming and Meng, Chenlin and Ermon, Stefano},
  title     = {Denoising Diffusion Implicit Models},
  booktitle = {International Conference on Learning Representations (ICLR)},
  year      = {2021}
}

@inproceedings{salimans2022progressive,
  author    = {Salimans, Tim and Ho, Jonathan},
  title     = {Progressive Distillation for Fast Sampling of Diffusion Models},
  booktitle = {International Conference on Learning Representations (ICLR)},
  year      = {2022}
}

@inproceedings{karras2022edm,
  author    = {Karras, Tero and Aittala, Miika and Aila, Timo and Laine, Samuli},
  title     = {Elucidating the Design Space of Diffusion-Based Generative Models},
  booktitle = {Advances in Neural Information Processing Systems (NeurIPS)},
  volume    = {35},
  pages     = {26565--26577},
  year      = {2022}
}

@inproceedings{lipman2023flowmatching,
  author    = {Lipman, Yaron and Chen, Ricky T. Q. and Ben-Hamu, Heli and Nickel, Maximilian and Le, Matt},
  title     = {Flow Matching for Generative Modeling},
  booktitle = {International Conference on Learning Representations (ICLR)},
  year      = {2023}
}

@inproceedings{liu2023rectifiedflow,
  author    = {Liu, Xingchao and Gong, Chengyue and Liu, Qiang},
  title     = {Flow Straight and Fast: Learning to Generate and Transfer Data with Rectified Flow},
  booktitle = {International Conference on Learning Representations (ICLR)},
  year      = {2023}
}

@article{wu2024cddm,
  author  = {Wu, Tong and Chen, Zhiyong and He, Dazhi and Qian, Liang and Xu, Yin and Tao, Meixia and Zhang, Wenjun},
  title   = {{CDDM}: Channel Denoising Diffusion Models for Wireless Semantic Communications},
  journal = {IEEE Transactions on Wireless Communications},
  volume  = {23},
  number  = {9},
  pages   = {11168--11183},
  year    = {2024},
  doi     = {10.1109/TWC.2024.3379244}
}

@inproceedings{jiang2024diffsc,
  author    = {Jiang, Zeyu and Liu, Xinyu and Yang, Guofeng and Li, Wenjie and Li, An and Wang, Guangyao},
  title     = {{DIFFSC}: Semantic Communication Framework With Enhanced Denoising Through Diffusion Probabilistic Models},
  booktitle = {ICASSP 2024 -- IEEE International Conference on Acoustics, Speech and Signal Processing (ICASSP)},
  pages     = {13071--13075},
  year      = {2024},
  doi       = {10.1109/ICASSP48485.2024.10448094}
}

@inproceedings{blau2018perception,
  author    = {Blau, Yochai and Michaeli, Tomer},
  title     = {The Perception-Distortion Tradeoff},
  booktitle = {IEEE/CVF Conference on Computer Vision and Pattern Recognition (CVPR)},
  pages     = {6228--6237},
  year      = {2018},
  doi       = {10.1109/CVPR.2018.00652}
}

@inproceedings{kingma2021vdm,
  author    = {Diederik P. Kingma and Tim Salimans and Ben Poole and Jonathan Ho},
  title     = {Variational Diffusion Models},
  booktitle = {Advances in Neural Information Processing Systems (NeurIPS)},
  volume    = {34},
  pages     = {21696--21707},
  year      = {2021}
}

@ARTICLE{10702555,
  author={Eldeeb, Eslam and Shehab, Mohammad and Alves, Hirley},
  journal={IEEE Wireless Communications Letters}, 
  title={A Multi-Task Oriented Semantic Communication Framework for Autonomous Vehicles}, 
  year={2024},
  volume={13},
  number={12},
  pages={3469-3473},
  doi={10.1109/LWC.2024.3472211}}

@article{shannon1948mathematical,
  author  = {Shannon, Claude E.},
  title   = {A Mathematical Theory of Communication},
  journal = {The Bell System Technical Journal},
  volume  = {27},
  number  = {3},
  pages   = {379--423},
  year    = {1948},
  doi     = {10.1002/j.1538-7305.1948.tb01338.x}
}

@article{gallager1962ldpc,
  author  = {Gallager, Robert G.},
  title   = {Low-Density Parity-Check Codes},
  journal = {IRE Transactions on Information Theory},
  volume  = {8},
  number  = {1},
  pages   = {21--28},
  year    = {1962},
  doi     = {10.1109/TIT.1962.1057683}
}

@misc{bellard2018bpg,
  author = {Bellard, Fabrice},
  title  = {{BPG} Image Format},
  year   = {2018},
  note   = {\url{https://bellard.org/bpg/}}
}

@book{proakis2008digital,
  author    = {Proakis, John G. and Salehi, Masoud},
  title     = {Digital Communications},
  edition   = {5th},
  publisher = {McGraw-Hill},
  year      = {2008}
}

@inproceedings{he2016resnet,
  author    = {He, Kaiming and Zhang, Xiangyu and Ren, Shaoqing and Sun, Jian},
  title     = {Deep Residual Learning for Image Recognition},
  booktitle = {IEEE Conference on Computer Vision and Pattern Recognition (CVPR)},
  pages     = {770--778},
  year      = {2016},
  doi       = {10.1109/CVPR.2016.90}
}

@inproceedings{ronneberger2015unet,
  author    = {Ronneberger, Olaf and Fischer, Philipp and Brox, Thomas},
  title     = {{U-Net}: Convolutional Networks for Biomedical Image Segmentation},
  booktitle = {Medical Image Computing and Computer-Assisted Intervention -- MICCAI 2015},
  pages     = {234--241},
  year      = {2015},
  publisher = {Springer},
  doi       = {10.1007/978-3-319-24574-4_28}
}

@inproceedings{balle2016gdn,
  author    = {Ball\'e, Johannes and Laparra, Valero and Simoncelli, Eero P.},
  title     = {Density Modeling of Images using a Generalized Normalization Transformation},
  booktitle = {International Conference on Learning Representations (ICLR)},
  year      = {2016}
}

@inproceedings{kingma2015adam,
  author    = {Kingma, Diederik P. and Ba, Jimmy},
  title     = {Adam: A Method for Stochastic Optimization},
  booktitle = {International Conference on Learning Representations (ICLR)},
  year      = {2015}
}

@article{wang2004ssim,
  author  = {Wang, Zhou and Bovik, Alan C. and Sheikh, Hamid R. and Simoncelli, Eero P.},
  title   = {Image Quality Assessment: From Error Visibility to Structural Similarity},
  journal = {IEEE Transactions on Image Processing},
  volume  = {13},
  number  = {4},
  pages   = {600--612},
  year    = {2004},
  doi     = {10.1109/TIP.2003.819861}
}

@inproceedings{wang2003msssim,
  author    = {Wang, Zhou and Simoncelli, Eero P. and Bovik, Alan C.},
  title     = {Multi-Scale Structural Similarity for Image Quality Assessment},
  booktitle = {Asilomar Conference on Signals, Systems \& Computers},
  volume    = {2},
  pages     = {1398--1402},
  year      = {2003},
  doi       = {10.1109/ACSSC.2003.1292216}
}

@inproceedings{zhang2018lpips,
  author    = {Zhang, Richard and Isola, Phillip and Efros, Alexei A. and Shechtman, Eli and Wang, Oliver},
  title     = {The Unreasonable Effectiveness of Deep Features as a Perceptual Metric},
  booktitle = {IEEE/CVF Conference on Computer Vision and Pattern Recognition (CVPR)},
  pages     = {586--595},
  year      = {2018},
  doi       = {10.1109/CVPR.2018.00068}
}

@inproceedings{heusel2017fid,
 author = {Heusel, Martin and Ramsauer, Hubert and Unterthiner, Thomas and Nessler, Bernhard and Hochreiter, Sepp},
 booktitle = {Advances in Neural Information Processing Systems},
 pages = {},
 publisher = {Curran Associates, Inc.},
 title = {{GANs} Trained by a Two Time-Scale Update Rule Converge to a Local Nash Equilibrium},
 volume = {30},
 year = {2017}
}

@inproceedings{cordts2016cityscapes,
  author    = {Cordts, Marius and Omran, Mohamed and Ramos, Sebastian and Rehfeld, Timo and Enzweiler, Markus and Benenson, Rodrigo and Franke, Uwe and Roth, Stefan and Schiele, Bernt},
  title     = {The Cityscapes Dataset for Semantic Urban Scene Understanding},
  booktitle = {IEEE Conference on Computer Vision and Pattern Recognition (CVPR)},
  pages     = {3213--3223},
  year      = {2016},
  doi       = {10.1109/CVPR.2016.350}
}

@inproceedings{ray2026hazematching,
  author    = {Ray, Anirban and Ashesh and Jug, Florian},
  title     = {HazeMatching: Dehazing Light Microscopy Images with Guided Conditional Flow Matching},
  booktitle = {Proc. IEEE/CVF Conf. Comput. Vis. Pattern Recognit. (CVPR) Findings},
  year      = {2026},
  doi       = {10.48550/arXiv.2506.22397},
}

@inproceedings{martin2025pnpflow,
  author    = {Martin, S{\'e}gol{\`e}ne and Gagneux, Anne and Hagemann, Paul and Steidl, Gabriele},
  title     = {PnP-Flow: Plug-and-Play Image Restoration with Flow Matching},
  booktitle = {Proc. Int. Conf. Learn. Represent. (ICLR)},
  year      = {2025},
  doi       = {10.48550/arXiv.2410.02423},
}

@inproceedings{krizhevsky2012alexnet,
  author    = {Krizhevsky, Alex and Sutskever, Ilya and Hinton, Geoffrey E.},
  title     = {ImageNet Classification with Deep Convolutional Neural Networks},
  booktitle = {Advances in Neural Information Processing Systems},
  volume    = {25},
  pages     = {1097--1105},
  year      = {2012}
}

@inproceedings{liu2023i2sb,
  author    = {Liu, Guan-Horng and Vahdat, Arash and Huang, De-An and Theodorou, Evangelos A. and Nie, Weili and Anandkumar, Anima},
  title     = {{I$^2$SB}: Image-to-Image Schr{\"o}dinger Bridge},
  booktitle = {Proc. Int. Conf. Mach. Learn. (ICML)},
  year      = {2023},
  doi       = {10.48550/arXiv.2302.05872},
}

@article{zhang2025semanticsguided,
  author  = {Zhang, Mingyu and Wu, Hongyu and Zhu, Guangyi and Jin, Rui and Chen, Xiaqing and G{\"u}nd{\"u}z, Deniz},
  title   = {Semantics-Guided Diffusion for Deep Joint Source-Channel Coding in Wireless Image Transmission},
  journal = {IEEE Trans. Wireless Commun.},
  year    = {2025},
  doi	  = {10.48550/arXiv.2501.01138}
}

@inproceedings{yilmaz2024perceptual,
  author    = {Yilmaz, Selim F. and Niu, Xueyan and Bai, Bo and Han, Wei and Deng, Lei and G{\"u}nd{\"u}z, Deniz},
  title     = {High Perceptual Quality Wireless Image Delivery with Denoising Diffusion Models},
  booktitle = {Proc. IEEE INFOCOM Workshops (DeepWireless Workshop)},
  year      = {2024},
  pages     = {345--350},
  doi 		= {10.48550/arXiv.2309.15889}
}

@ARTICLE{10948463,
  author={Eldeeb, Eslam and Shehab, Mohammad and Alves, Hirley and Alouini, Mohamed-Slim},
  journal={IEEE Transactions on Machine Learning in Communications and Networking}, 
  title={Semantic Meta-Split Learning: A {TinyML} Scheme for Few-Shot Wireless Image Classification}, 
  year={2025},
  volume={3},
  number={},
  pages={491-501},
  doi={10.1109/TMLCN.2025.3557734}}

@ARTICLE{11604027,
  author={Fu, Jingwen and Xiao, Ming and Skoglund, Mikael and Kim, Dong In},
  journal={IEEE Transactions on Wireless Communications}, 
  title={Land-Then-Transport: A Flow Matching-Based Generative Decoder for Wireless Image Transmission}, 
  year={2026},
  volume={25},
  number={},
  pages={19757-19772},
  doi={10.1109/TWC.2026.3710439}}

@misc{gao2026lowlatencygenerativesemanticcommunication,
      title={Low-Latency Generative Semantic Communication via Channel-Realization Flow Matching}, 
      author={Fan Gao and Youzheng Wang and Zhijin Qin and Feifei Gao},
      year={2026},
      eprint={2607.24876},
      archivePrefix={arXiv},
      primaryClass={cs.IT},
      url={https://arxiv.org/abs/2607.24876}, 
}

\end{document}